\documentclass[superscriptaddress,groupedaddress,nofootnoteinbib,11pt]{article}
\pdfoutput=1 
\usepackage{jheppub}

\usepackage[utf8x]{inputenc}
\usepackage{mathtools,slashed,mathrsfs}
\usepackage[caption=false]{subfig}
\usepackage{dcolumn}
\usepackage{multirow}
\usepackage{tabularx}
\usepackage{float}
\usepackage{booktabs}
\usepackage{bm}
\usepackage{comment}
\usepackage{setspace}
\usepackage[dvipsnames]{xcolor}
\usepackage[normalem]{ulem} 
\usepackage{enumerate}
\usepackage{siunitx}
\graphicspath{{../Mathematica/}}

\def\AA{\mathcal{A}}

\def\mat{\begin{pmatrix}}
\def\rix{\end{pmatrix}}
\def\ket{\right \rangle}
\def\bra{\left\langle}
\def\rar{\rightarrow}

\def\({\left(}
\def\){\right)}
\def\[{\left[}
\def\]{\right]}

\def\avg{\overline}

\def\where{\quad \text{where}\quad}

\def\vec{\mathbf}
\def\vec{\boldsymbol}
\def\MM{\mathcal{M}}
\def\OO{\mathcal{O}}

\newcommand\bea{\begin{equation}}
\newcommand\eea{\end{equation}}

\begin{document}

\title{Supernova constraints on an axion-photon-dark photon interaction}

\date{\today}
\author[a]{Anson Hook,}
\author[a]{Gustavo Marques-Tavares,}
\author[a]{Clayton Ristow}

\affiliation[a]{Maryland Center for Fundamental Physics, University of Maryland, College Park, MD 20742, U.S.A.}

\emailAdd{hook@umd.edu}
\emailAdd{gusmt@umd.edu}
\emailAdd{cristow@umd.edu}

\abstract{
We present the supernova constraints on an axion-photon-dark photon coupling, which can be the leading coupling to dark sector models and can also lead to dramatic changes to axion cosmology. We show that the supernova bound on this coupling has two unusual features.
One occurs because the scattering that leads to the trapping regime
converts axions and dark photons into each other.
Thus, if one of the two new particles is sufficiently massive, both production and scattering become suppressed and the bounds from bulk emission and trapped (area) emission both weaken exponentially and do not intersect. 
The other unusual feature occurs because for light dark photons, longitudinal modes couple more weakly than transverse modes do.  Since the longitudinal mode is more weakly coupled, it can still cause excessive cooling even if the transverse mode is trapped.
Thus, the supernova constraints for massive dark photons look like two independent supernova bounds super-imposed on top of each other.
}

\maketitle

\section{Introduction}

One of the most extreme environments in the universe are the centers of core-collapse supernovae. For many seconds right after the collapse, supernovae reach nuclear densities and temperatures of order 10s of MeV before they are cooled via neutrino emission. This makes them a potential source for particles beyond the Standard Model (BSM) if their masses are below about 100 MeV.

The detection of neutrinos from the supernova SN1987A~\cite{Hirata:1987hu,Bionta:1987qt,Alekseev:1987ej} confirmed the theoretical picture that the proto-neutron stars formed during core collapse supernova cool via neutrino emission.  Thus, one can place bounds on BSM particles by requiring that they cannot be emitted so much as to cool the supernovae faster than neutrinos~\cite{Burrows:1986me,Burrows:1987zz,Raffelt:1996wa}. This allows one to constrain particles that are so weakly coupled that they can escape the core without interacting, which results in a bulk emission constraint. For larger couplings, the new particles become trapped and the cooling is effectively only done by particles emitted at the last scattering surface, which acts like a blackbody of the trapped particles. The trapping regime leads to an upper bound on the couplings that can be constrained by cooling while bulk emission leads to a lower bound.  Generically, the lower bound curve from bulk emission and the upper bound curve from surface emission in the mass vs coupling plane intersect at large masses, leading to a closed region that can be excluded by cooling considerations.

Since SN1987A, cooling arguments have been used to constrain axions~\cite{Turner:1987by,Raffelt:1987yt,Raffelt:1988py,Lucente:2020whw}, CP even scalars~\cite{Ishizuka:1989ts}, dark photons~\cite{Bjorken:2009mm, Dent:2012mx, Rrapaj:2015wgs}, dark sectors~\cite{Dreiner:2013mua,Chang:2018rso,Camalich:2020wac}, sterile neutrinos~\cite{Kainulainen:1990bn}, extra dimensions~\cite{Hanhart:2000er}, and supersymmetry~\cite{Hanhart:2001fx}.  In the process of deriving these bounds, 
it was slowly realized that many important effects and features were not initially properly accounted for.  For example, finite density effects drastically changed the qualitative features of supernovae bounds~\cite{Chang:2016ntp,Hardy:2016kme}.  We will continue this trend by pointing out two new features present in supernova bounds.

In this paper, we study the supernovae cooling constraint on the axion-photon-dark-photon coupling, and show that this model leads to two new qualitative features in supernovae that can play a role in non-minimal scenarios where there are more than one new degree of freedom.  The first feature we discuss
is that different polarizations of particles can behave differently,
and thus their production, and in particular their trapping, needs to be treated separately rather than averaged over\footnote{The fact that the production rate of transverse and longitudinal modes have to be treated separately has long been recognized by many previous references.  Here we are emphasizing the importance of treating distinct polarizations differently in the trapping regime as well.}.
An example of where something like this occurs is when discussing
transverse and longitudinal modes of a dark photon.  The longitudinal mode
couples more weakly than the transverse mode and needs its own
cooling and trapping constraints.

The second feature arises when the new particles must be pair produced and the two species have different masses.  
For fixed coupling, as the mass of one particle becomes larger,
their joint production is suppressed because the threshold energy for pair production is increased.  However, because the same interaction vertex appears in scattering
processes, the scattering of the lighter particle becomes suppressed as well, as it does not have enough energy to up-scatter into the heavier particle.  As a result, both production and scattering become suppressed so that the associated bounds both become weaker exponentially, and the trapping limit and bulk emission limit curves do not intersect.

In the axion-photon-dark-photon model we study in this paper, there are two new particles, the axion and the dark photon, and their leading coupling is given by
\bea
\label{eq: coupling}
\mathcal{L} \supset \frac{1}{2} \frac{a}{f_a} F_{\mu \nu} \tilde F^{\mu \nu}_D ,
\eea
where $a$ is the axion and $F$ ($F_D$) is the field strength tensor of the photon (dark photon).  The axion and the dark
photon are two of the most motivated candidates for physics beyond the standard model.
Axions are 
pseudo-Goldstone bosons that appear very naturally in string theory~\cite{Arvanitaki:2009fg}
and can solve a variety of problems such as the strong CP problem~\cite{Peccei:1977np,Peccei:1977hh,Weinberg:1977ma,Wilczek:1977pj}, or
dark matter~\cite{Abbott:1982af,Dine:1982ah,Preskill:1982cy}.  Dark photons appear in a wide range of
motivated models of dark matter~\cite{Nelson:2011sf,Arias:2012az,Battaglieri:2017aum} or as a solution to various
anomalies~\cite{Gninenko:2001hx,Pospelov:2008zw,TuckerSmith:2010ra,Batell:2011qq,Altmannshofer:2017bsz}. While in many models the leading coupling of the dark photon to our sector is through kinetic mixing, this coupling can be highly suppressed making the interaction in Eq.~\ref{eq: coupling} the most relevant one for phenomenology.  In particular, if the dark sector has a charge conjugation symmetry, $C_D$, then kinetic mixing is removed.  In this case, as long as the axion is odd under both $C_D$ and $CP$,  Eq.~\ref{eq: coupling} is the leading coupling between the dark sector and the visible sector.

A coupling of the form shown in Eq.~\ref{eq: coupling} has been considered before in many different contexts, e.g. Ref.~\cite{Kaneta:2016wvf,Kaneta:2017wfh,Pospelov:2018kdh,Choi:2018mvk,Kalashev:2018bra,Biswas:2019lcp,Choi:2019jwx,Hook:2019hdk,deNiverville:2020qoo,Arias:2020tzl}.  Additional motivation for studying this coupling comes from dark photon dark matter.  Dark photons are a compelling dark matter candidate and so production mechanisms for it are interesting.  Above a keV, dark photon dark matter can be produced thermally, either through freeze-out or freeze-in.  As it becomes lighter than keV, one needs to consider non-thermal production mechanisms~\cite{Graham:2015rva,Agrawal:2018vin,Bastero-Gil:2018uel,Co:2018lka,Dror:2018pdh,Long:2019lwl}.  Dark photon production from axions using the coupling in Eq.~\ref{eq: coupling} is one of the few ways to realize dark photon dark matter lighter than a $\mu$eV~\cite{longpaper}. This coupling also leads to the two new aforementioned features on the cooling bound, it has a dark photon whose longitudinal mode and transverse modes
couple to matter with different strengths.  Additionally, because the coupling Eq.~\ref{eq: coupling} contains both the dark photon and the axion, production and scattering necessarily
involve two separate species, which must be taken into account in cooling constraints.  The bounds on this coupling are shown in Fig.~\ref{Fig: Full Constraints} where both of these features can be seen.  At low mass, there are two separate cooling and trapping constraints while at high mass, both the trapping and cooling bounds become exponentially weaker.

In Sec.~\ref{Sec: SN Emission}, we review how core-collapse supernovae can be used to place constraints on new physics and derive the cooling constraints on our model.
In Sec.~\ref{Sec: Trapping}, we derive the trapping constraints.
We discuss in detail the longitudinal mode of the dark photon and its constraints in Sec.~\ref{Sec: Longitudinal}.
Non-supernova constraints are discussed in Sec.~\ref{Sec: WD and BBN}.
Finally, in Sec.~\ref{Sec: conclusion} we conclude.

\section{The Energy Loss Argument and Emission Constraints}
\label{Sec: SN Emission}

The core collapse supernova SN1987A has provided, and continues to provide, valuable constraints on new physics. Measurements of the energy emitted into neutrinos and the duration of the emission were used to infer the cooling rate of the supernova. This cooling rate was found to be consistent with Standard Model predictions. Any new particle weakly coupled to the Standard Model provides a new channel for energy loss in the supernova. This would increase the cooling rate, pushing it out of agreement with Standard Model predictions, and provides a constraint on the coupling of new light particles to the Standard Model. This constraint can be expressed in a simple form with the Raffelt criterion~\cite{Raffelt:1996wa}, which states that the luminosity of new particles, $L_x$, can be no larger than the luminosity of neutrinos, $L_\nu$,
\bea
L_x\leq L_\nu =3 \cdot 10^{52} \; \text{erg$\cdot$s$^{-1}$} \, .
\label{Eq: Raffelt Criterion}
\eea
Depending on how strongly the new particles interact, cooling can occur in two different regimes. If the new particles interact very weakly, after being produced they carry away energy without further interactions inside the proto-neutron star. This corresponds to the bulk emission regime and is used to place a lower bound on the coupling, dependent only on the amount of energy going into the new particles. Once the interactions of the new particles are sufficiently strong, they become trapped in the proto-neutron star and the cooling is only done through emission at their last scattering surface (trapping surface), i.e. as in a blackbody. As the coupling becomes stronger, this trapping surface moves to larger radii and thus to lower temperatures, which results in a smaller luminosity. This leads to an upper bound on the couplings that can be excluded by Raffelt's criterion. In this section we will examine the constraint arising from bulk emission and discuss the trapping constraint in Sec.~\ref{Sec: Trapping}.

In our model the axion and dark photon play the roles of these weakly interacting particles. To find the luminosity of axions and dark photons we must find the emissivity, $\dot \epsilon$, or energy emission rate per unit volume. Integrating this over the volume of the supernova gives the luminosity of the bulk emission,
\begin{equation*}
L_{Bulk}=\int_{V_{SN}} d^3\vec r \;\dot \epsilon(r) \, . 
\end{equation*}

Next, we must compute the emissivity for each process that can produce axions and dark photons in the supernova. To find these emissivites, we simply take the interaction rate per unit volume for that process and weigh it by the energy carried away by the axion and dark photon in the final state~\cite{Dreiner:2013mua},
\bea
\dot \epsilon =\(\prod_i \int\frac{d^3 \vec p_i}{(2\pi)^3 2 E_i} f_i(E_i)\)\(\prod_f \int\frac{d^3 \vec p_f}{(2\pi)^3 2 E_f}\)\avg{|\MM|^2}(E_a+E_D)(2\pi)^4\delta^4\(\sum p_i-\sum p_f\) \, .
\label{Eq: Emissivity General}
\eea
Here, momenta indexed with $i$ represent initial state particles while momenta indexed with $f$ represent final state particles.  $f_i$ represent the thermal weights of the initial state particles. The matrix element is averaged over initial spins and summed over final spins. $E_a$ and $E_D$ are the energies of the produced axion and dark photon. For computational ease, we neglect Pauli blocking and Bose enhancement of the final states. The latter is never a large effect in the processes under consideration, while the former can significantly reduce the rates for processes involving scattering with electrons at the core. As we will show later, those processes are subdominant even without taking Pauli blocking into account, and thus we will not include them when deriving constraints. When Eq.~\ref{Eq: Emissivity General} is simplified for processes involving two initial state particles ($A$ and $B$), the emissivity can be nicely written as the product of initial densities multiplied by a thermally averaged, energy weighted cross section
\bea
\dot{\epsilon}_{A+B\rar a+\gamma_D+X}=n_An_B\bra v_{mol} \sigma_E\ket \, , 
\label{Eq: Emissivities 2 to N}
\eea
where $\sigma_E$ is the ordinary cross section weighted by the sum of energies of the axion and dark photon. The brackets indicate an averaging over initial thermal states in the supernova and is defined by Eq.~\ref{Eq: Thermal Average} in App.~\ref{App: Computations}. The integral over final states, $\int d\Pi_f$ is defined in App.~\ref{App: Computations} in Eq.~\ref{Eq: Final State Integral}, while $v_{mol}$ is the frame invariant Moller velocity, $v_{mol}=\sqrt{(\vec v_A- \vec v_B)^2+(\vec v_A\times \vec v_B)^2}$.

A similar simplification can be done for the emissivity for any process with a single initial state particle ($A$). The result is a product of the initial densities and the average decay rate weighted by energy is
\bea
\dot \epsilon_{A\rar a+\gamma_D}=n_A\bra \Gamma_{A\rar a+\gamma_D} E_A \ket ,
\label{Eq: Emissivities Decay}
\eea
where $\Gamma$ is the usual decay rate.

\begin{figure}[H]
\centering
\includegraphics[width=.9\linewidth]{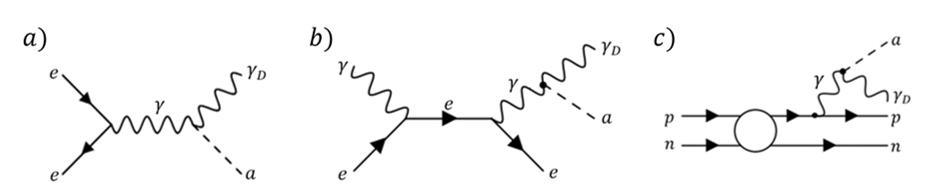}
\caption{Representative figures for the 3 leading collision processes that produce axions and dark photons in a supernova. a) $e^+e^-\rar \gamma_D a$, the annihilation of an electron and positron.
b) \indent  $e^+\gamma\rar e^+\gamma_D a$,  Compton scattering where the off-shell Standard Model photon decays to an axion and dark photon. c) $pn\rar pn\gamma_D a$, nuclear bremsstrahlung where the off-shell photon decays to an axion and dark photon.}
\label{Fig: 2 to N Processes}
\end{figure}

There are 4 relevant processes for the production of axions and dark photons in supernova. 3 of them are collision processes involving 2 initial state particles: the annihilation of an electron and positron, Compton scattering of an electron and photon, and nuclear bremsstrahlung of a proton and neutron. An example diagram for each of these processes is shown in Fig.~\ref{Fig: 2 to N Processes}.  

The emissivites for these processes are computed in App.~\ref{App: Computations} and the results are given in Eq.~\ref{Eq: Computed Emissivities 2 to N},
\bea
\begin{aligned}
\dot \epsilon_a =\frac{\alpha_e} {24 f_a^2}n_{e^-}n_{e^+}\bra E_{tot}\ket Q^a(m_a,m_D) \, , \\
\dot \epsilon_b=\frac{\alpha_eT^3}{90 \pi^3 f_a^2}n_pn_n\bra \Sigma_{pn}\ket Q^b(m_a,m_D) \, , \\
\dot \epsilon_c=\frac{\alpha_e^2}{48 \pi f_a^2}  n_{e^-}n_\gamma\bra E_{tot}\ket Q^c(m_a,m_D) \, , 
\end{aligned}	
\label{Eq: Computed Emissivities 2 to N}
\eea
where $\langle E_{tot} \rangle$ is the average of the sum of the initial state particle's energies. The $Q$ factors in each expression, defined in Eqs.~\ref{Eq: Qa}, \ref{Eq: Qb}, \ref{Eq: Qc} in App.~\ref{App: Computations}, contain all of the dependence on the masses of the axion and dark photon. They are normalized so that they are $\OO(1)$ when $m_a=m_D=0$ and then decrease as the axion or dark photon gain mass.

The annihilation and Compton emissivities have a simple structure. They include the couplings involved in the process, the initial densities and the average of the total initial energy. The bremsstrahlung emissivity has a different structure due to how it was computed, which we do following Ref.~\cite{Chang:2018rso}. In order to deal with the nuclear scattering, which does not allow for a simple perturbative calculation in the energy regime of interest, we assume that the process is dominated by scatterings where the radiated particles (and hence the virtual photon in the diagram) have energies much smaller than the kinetic energy of the nucleons~\footnote{Cooling is dominated by particles with energies comparable to the temperature at which they are produced, and thus there is a large uncertainty associated with this approximation. In Ref.~\cite{Rrapaj:2015wgs} it is argued that for real photon emissions with energies comparable to the nucleon energies this approximation is accurate to within a factor of 2.
}. 
This approximation, called the soft radiation approximation, allows us to factorize the emissivity into two factors: one factor accounting for the nuclear scattering and another capturing the decay of a virtual photon into the axion and dark photon. The factor $\Sigma_{pn}$ is the piece that accounts for the dynamics of the nuclear scattering and is proportional to the momentum transfer cross section of proton-neutron scattering.  In detail, it is given by
\bea
\Sigma_{pn}=\frac{|\vec p_p|^2}{m_N^2}v_{mol} \sigma^{pn}_{\Delta \vec p} \where \sigma^{pn}_{\Delta \vec p}=\int d\Omega (1-\cos(\theta))\frac{d\sigma_{pn}}{d\Omega} .
\label{Eq: Nuclear Cross Section}
\eea
In order to compute the momentum transfer cross section we use experimental data from Ref.~\cite{Rrapaj:2015wgs} and for more details see App.~\ref{App: Computations}.

\begin{figure}[H]
\centering
\includegraphics[width=.25\linewidth]{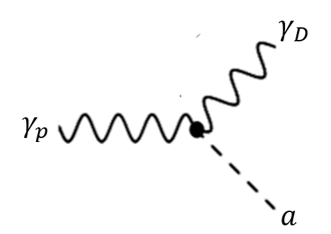}
\caption{The leading order decay process that can produce a dark photon and axion. A plasmon $\gamma_p$ decays into an axion and a dark photon.}
\label{Fig: Decay Process}
\end{figure}

Now we consider the decay processes that could produce axions and dark photons. To leading order, there is only one such process: the decay of plasmons into dark photons and axions. The diagram of this process is shown in Fig.~\ref{Fig: Decay Process}.

In order to compute the emissivity, we've used the polarizations and dispersion relations for the plasmon given in Ref.~\cite{Braaten:1993jw}. Again, the details of the computation are given in App.~\ref{App: Computations}, but the result is 
\bea
\dot \epsilon_p=\frac{\zeta(3)T^3}{3\pi f_a^2}\(\frac{\omega_p^2}{4\pi}\)^2Q^p(m_a,m_D),
\label{Eq: Computed Emissivity Decay}
\eea
where $Q^p$ defined in Eq.~\ref{Eq: Qp} contains all of the dependence on the mass of the axion and dark photon. It is $\OO(1)$ when $m_a=m_D=0$ and decreases for increasing dark photon and axion masses. $\omega_p$ is the plasmon frequency  defined in Eq. ~\ref{Eq: Plasmon Frequency}. 

\begin{figure}[t]
\centering
\includegraphics[width=.8\linewidth]{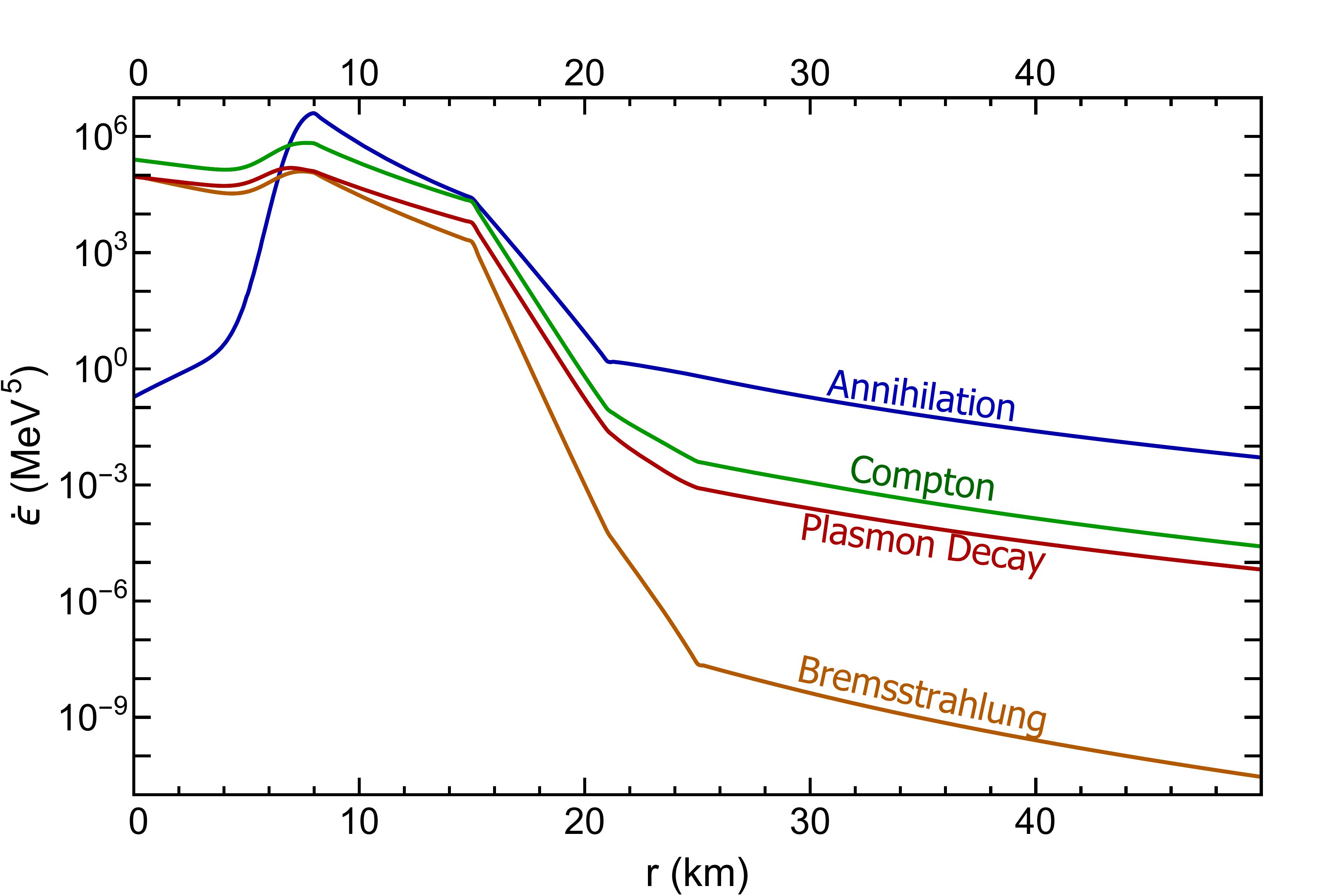}
\caption{A plot of the emissivities of the various processes that produce axions and dark photons as a function of radius of the supernova. These emissivities are taken for both the dark photon and axion massless. The annihilation process is the dominant process everywhere in the supernova except for near the center of the supernova where it is suppressed by the low density of positrons. The coupling has been set to $f_a=1$ MeV for convenience.}
\label{Fig: Emissivity Profiles}
\end{figure}

Fig.~\ref{Fig: Emissivity Profiles} shows a plot of the emissivities for $m_a=m_D=0$ as a function of supernova radius. These emissivities are computed using the temperature and density profiles given in App.~\ref{App: Profiles}. Near the center of the core, Compton scattering and plasmon decay dominate the energy emission while at larger radii, annihilation dominates.  Note that the Compton emissivity does not take Pauli blocking into account, and thus the true rate would have a large suppression compared to what is shown in Fig.~\ref{Fig: Emissivity Profiles}. We will not include the Compton contribution to the luminosity, which is a conservative approach.  Since the Compton contribution to the total luminosity is subdominant, neglecting it only leads to a small effect. Upon integration of each emissivity over the volume of the supernova, we find that the annihilation is the dominant process, with plasmon decay being a significant contribution for lower masses.

Finally, these emissivities are summed into a total emissivity and integrated over the volume of the supernova to give the total luminosity of axions and dark photons. This total luminosity is then applied to the constraint in Eq.~\ref{Eq: Raffelt Criterion} for various axion and dark photon masses and is shown in the lower line in Fig.~\ref{Fig: Supernova Bounds}.

\section{Trapping Constraints} 
\label{Sec: Trapping}

As discussed in Sec.~\ref{Sec: SN Emission}, at stronger couplings cooling is done through surface emission due to particles becoming trapped. Using this surface emission, one can derive an upper bound on the couplings from the Raffelt criterion in Eq.~\ref{Eq: Raffelt Criterion}. Once the coupling is large enough, any new particle, $x$, will be trapped inside the supernova below some radius $r_x$. This radius, $r_x$, is defined as the point where the optical depth $\tau$ is equal to $2/3$
\bea
\frac{2}{3}=\tau(r_x) . 
\label{Eq: Optical Depth Definition}
\eea
The optical depth is defined as
$$\tau(r_x)=\int_{r_x}^\infty \frac{dr}{\lambda_x} ,$$
where $\lambda_x$ is the mean free path of the new particle and is related to the scattering process for the new particle $x$ with Standard Model particles.  Because of rapid absorption and re-emission, the emitted energy due to this new particle is no longer the bulk emission seen in Sec.~\ref{Sec: SN Emission} but can be approximated as blackbody radiation from this radius\footnote{
In general there are separate radii at which number changing processes, energy exchange	processes and trapping freeze-out and this can be taken into account in order to improve the bound as discussed in Refs.~\cite{Raffelt:2001kv,DeRocco:2019jti}. Here we take the simplified approach of considering the trapping radii, which is the largest of the three, as the blackbody since this leads to a smaller flux and hence a more conservative limit.
} $r_x$ given by
\bea
L_{Blackbody}(r_x,m_x,g_{\star,x})=4 \pi r_x^2\(\frac{g_{\star,x} \pi^2}{120}\) T^4(r_x) h(m_x/T) .
\label{Eq: Blackbody Radiation}
\eea
Here $m_x$ is the mass of the particle and $g_{\star,x}$ is its number of degrees of freedom. This resembles the familiar expression for blackbody radiation with $\frac{g_{\star,x} \pi^2}{120}$ being the Steffan Boltzmann constant in natural units and an additional factor $h(m_x/T)$. This factor is to account for the mass of the radiated particle and is defined by
\bea
h(z)=\frac{15}{\pi^4}\int_{z}^{\infty}dx\frac{x(x^2-z^2)}{e^x-1} .
\label{Blackbody Mass Suppression}
\eea
This integral is such that when the particle is massless, $h(0)=1$ and we recover the familiar blackbody formula. In the limit $m_x \gg T$, $h \sim e^{-m_x/T}$ meaning that for very massive particles, blackbody radiation is highly suppressed.

The luminosity of this blackbody radiation has a different dependence on the coupling than the bulk emission considered in Sec.~\ref{Sec: SN Emission}. Since for radii larger than the core radius, $\sim 10$ km, $T$ falls off faster than $r^{-1/2}$ (see App.~\ref{App: Profiles}), the total luminosity decreases with increasing $r$. This means that as the new particle becomes more and more weakly coupled, it can travel through the star easier and so the radius of blackbody emission shrinks.  As the blackbody radius shrinks, the emitted luminosity increases (assuming it is outside the proto-neutron star's core) and so in the trapping limit, the Raffelt criterion excludes all couplings {\it weaker} than a certain value.

Now we will apply this trapping constraint to our model. Both the axion and dark photon have their own mean free paths $\lambda_a$ and $\lambda_D$ from their decay and scattering with SM particles. From these mean free paths we define the two radii $r_a$ and $r_D$ via Eq.~\ref{Eq: Optical Depth Definition}. Effectively, these radii can be thought of as the radius below which the corresponding particle is trapped and outside of which it free-streams. 

The difference between $r_a$ and $r_D$ introduces extra complications that precludes a simple treatment of the luminosity of both dark photons and axions as arising from a blackbody. This arises from the fact that in the region between the two radii neither of the particles are in thermal equilibrium with the rest of the proto-neutron star, and so the assumptions that go into treating the outer radius as a blackbody fail. To illustrate these complications, let us consider the case where $r_D<r_a$, i.e. the dark photon becomes free-streaming at a smaller radii than the axion.  While the dark photon can be treated as a blackbody emitting from $r_D$, the axion emission is not so simple. In the region between $r_D$ and $r_a$, the axion scatters or decays before traveling a large distance, but the dark photon can easily escape the region without subsequent interactions. Because any interaction/decay of the axion converts it into a dark photon, even though from Eq.~\ref{Eq: Optical Depth Definition} we find that the axion is trapped, after a single interaction a portion of its energy goes into dark photons that can escape the region. Therefore, we can treat the smaller radii as a blackbody for both particles, but we must take into account the transmission probability for the energy that is radiated into axions.

Once produced at the inner trapping radius, the particle associated with the larger trapping radius travels a finite distance before having an interaction (scattering or decay) that converts it into the free-streaming particle. The luminosity associated with the trapped particle depends on the probability that the free-streaming particle created in that interaction travels outward (so that it does not return to the inner radius), $P_{\theta}$, and the fraction of the initial particle energy carried by the free-streaming particle, $\epsilon$. Calculating those factors precisely is computationally intensive and amounts to small changes in the final result. To simplify, we take $P_\theta$ and $\epsilon$ to have the same value for all scatterings. So that the total luminosity is
\bea
L_{tot}( r_{in})=L_{Blackbody}(r_{in},m_{in},g_{\star,in})+\epsilon P_\theta L_{Blackbody}(r_{in},m_{out},g_{\star,out}) ,
\label{Eq: Modified Blackbody Radiation}
\eea
where $m_{in}, m_{out}, g_{\star,in},$ and $g_{\star,out}$ are the masses and degrees of freedom of the particle corresponding to the inner and outer radius respectively.  In reality, $P_\theta$ depends on how far the initial particle makes it from the blackbody before scattering since the further it makes it out, the more solid angle it can scatter into without being directed back into the blackbody. Thus we can say that $1/2\leq P_\theta\leq 1$. To get an estimate for $\epsilon$ we can easily see that depending on the mass of the initial particle and whether the event is a scattering or a decay that in the center of mass frame the final particle takes between $1/2$ and all of the initial particle's energy. Thus we have $1/2 \leq\epsilon\leq 1$. In order to get a conservative estimate on our coupling we note that if we underestimate our total luminosity, we will also underestimate the radius of the blackbody once we solve $L_{tot}( r_{in})=L_{\nu}$. This means we will be underestimating the constraint on the coupling thus giving us a conservative upper bound. We therefore take the conservative estimate of $\epsilon= P_\theta=1/2$. In order to get a sense of the maximum value our constraint could be off by, we look at the other extreme $\epsilon= P_\theta=1$.  This changes the value of the constraint curve by at most around $20\%$, so we use the value $\epsilon P_\theta=1/4$ as it provides a good conservative estimate of the constraint.

\begin{figure}[H]
\centering
\includegraphics[width=.55\linewidth]{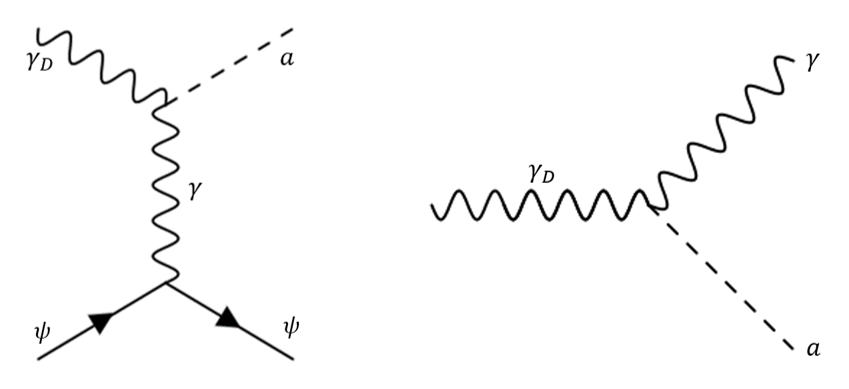}
\caption{The diagrams for the dark photon's scattering off of fermions and its decay into an axion and photon. $\psi$ represents any of the electrically charged fermions in the supernova: electrons, positrons, and protons. The diagrams for axion's scattering from fermions and its decay are identical to this but with the axion and dark photon exchanged. These diagram are used to compute the momentum transfer cross sections for the various scattering processes and the decay width for the axion and dark photon. These cross sections are then used to place a trapping constraint on the coupling $1/f_a$.}
\label{Fig: Scattering Processes}
\end{figure}

Now we compute the mean free paths of the axion and dark photons. The mean free path for both particles is determined by their scattering cross section with the three electrically charged fermions in the supernova: electrons, positrons, and protons as well as their decay width (where only the heavier between the axion and dark photon decays). The diagrams for these scatterings and decays are shown in Fig.~\ref{Fig: Scattering Processes} with $\psi$ designating any of the three fermions. 

In order to compute the mean free path, we must compute the cross sections and decay widths from these interactions. However, rather than using the ordinary cross sections to determine the mean free paths, we will use the momentum transfer cross section defined as
\bea 
\sigma_{\Delta \vec p} =\frac{1}{4E_AE_B\Delta \vec v}\int d\Pi_f (1-\cos(\theta))|\MM|^2 .
\label{Eq: Momentum Transfer Cross Section}
\eea
Using the momentum transfer cross section means that we are favoring scatterings with a more severe angular deflection because they are more effective at trapping particles than those with a small angular deflection. The momentum transfer cross section also helps to regulate the $t$ channel divergence that appears in our diagrams when $m_a=m_D$. The computation of these cross sections is detailed in App.~\ref {App: Computations}, where we show the cross-sections take the form
\bea
\sigma_{a\psi\rar \gamma_D \psi}=\frac{\alpha_e}{3f_a^2}\frac{s\beta(s,m_\psi,m_D)}{2E_AE_B|\vec v_A-\vec v_B|}I(s,m_\psi^2,m_a^2,m_D^2) 
\label{Eq: Scattering Cross Sections}
\eea
$$\sigma_{\gamma_D\psi\rar a \psi}=\frac{\alpha_e}{3g_{\star, D}f_a^2}\frac{s\beta(s,m_\psi,m_a)}{2E_AE_B|\vec v_A-\vec v_B|}I(s,m_\psi^2,m_D^2,m_a^2),$$
where $m_\psi$ is the mass of the fermion the axion or dark photon is scattering with. The definition for $I$ is given Eq.~\ref{Eq: I Defintion} in App.~\ref{App: Computations}, and is exactly 1 for $m_a=m_D=0$. The decay widths are computed in App.~\ref{App: Computations} and are
\bea 
\Gamma_a=\frac{m_a^4}{32 \pi f_a^2 E}\(1-\frac{m_D^2}{m_a^2}\)^3 \, , \quad \quad \Gamma_D=\frac{m_D^4}{96\pi f_a^2 E}\(1-\frac{m_a^2}{m_D^2}\)^3,
\label{Eq: Decay Widths}
\eea
where $E$ is the energy of the initial particle. 
The mean free path $\lambda$ for the axion and dark photon are defined as 
\bea
\frac{1}{\lambda_a}=\Gamma_a+\sum_{\psi} n_\psi\bra  \sigma_{a,\psi}\ket_\psi \, , \quad \quad\frac{1}{\lambda_D}=\Gamma_D+\sum_{\psi} n_\psi\bra\sigma_{D,\psi}\ket_\psi \, ,
\label{Eq: Mean Free Path} 
\eea
where in the sum $\psi$ represents: electrons, positrons and protons. $\sigma_{a,\psi}$ is a short hand notation for the momentum transfer cross section of an axion with fermion $\psi$ and likewise $\sigma_{D,\psi}$ for the dark photon. Additionally, the subscript $\psi$ in the thermal average in Eq.~\ref{Eq: Mean Free Path} is meant to convey that we are taking the average of initial fermion states only. For the initial boson (axion or dark photon) energy we simply input the average energy of a boson at radius $r_i$.

It is important to highlight why inserting the average energy for the initial boson (call it $X$), rather than averaging over initial energies is the more physically accurate thing to do. This stems from the fact that particles of different energies decouple at different radii, and thus treating this radiation as a blackbody emission is an approximation. This approximation can become very inaccurate when applied to particles that have inelastic scatterings. Consider the scenario where the final state boson has a very large mass, $M$. 
In order to have enough energy to produce the much more massive final state boson, the particle $\psi$ that $X$ scatters off of needs to have very high energy.  The lower the energy of $X$, the higher the energy of $\psi$ needs to be.
This implies that $X$ with a lower energy will scatter less frequently than $X$ of higher energy. In the limit that $ M \gg T$, most of the initial state bosons are effectively free while only those with very high energy have a chance at scattering.  Averaging over initial state $X$ energies fails to capture this behavior effectively because it will assign every $X$ the same mean free path regardless of energy. 
This would lead us to conclude that the typical thermal bosons in the supernova scatter much more effectively than they actually do which in turn would lead us to overestimate the extent to which these particles are trapped in the supernova.
In order to capture the dynamics, which is dominated by particles with energies close to the average energy, it is more physically accurate to use the mean free path for particles with energy $\bra E_b\ket$. So, the thermal average used in Eq.~\ref{Eq: Mean Free Path} is defined as
\bea
\bra \sigma_{a,\psi} \ket_\psi =\int \frac{d^3 \vec p_\psi f(E_\psi,T(r),\mu_\psi(r))}{(2\pi)^3n_\psi(r)} \sigma_{a,\psi}(E_\psi,\bra E_b \ket_{r_i}) ,
\label{Eq: Averaged Cross Section}
\eea
where
$$\bra E_b \ket_{r_i} =\int \frac{d^3 \vec  p_bf(E_b,T(r_i))}{(2\pi)^3n_b(r_i)}E_b.$$

Now we can use these cross sections and the corresponding mean free paths to place the constraint on the coupling. The radii  $r_a$ and $r_D$, are implicitly defined as functions of the coupling using Eq.~\ref{Eq: Optical Depth Definition} and using the mean free paths defined in Eq.~\ref{Eq: Mean Free Path}. Then we can numerically solve for the value of $f_a$ such that the emitted luminosity given in Eq.~\ref{Eq: Modified Blackbody Radiation} is equal to $L_\nu$
\bea
L_{tot}(\min(r_a(f_{a}),r_D(f_{a})))=L_\nu
\label{2 Particle Constraint}
\eea
Since we expect both radii to decrease with increasing $f_a$, and that the blackbody luminosity should increase as the radii decrease, we can exclude couplings below $1/f_{a}$ (but larger than the constraint from bulk emission).

\begin{figure}[t]
\centering
\includegraphics[width=.49\linewidth]{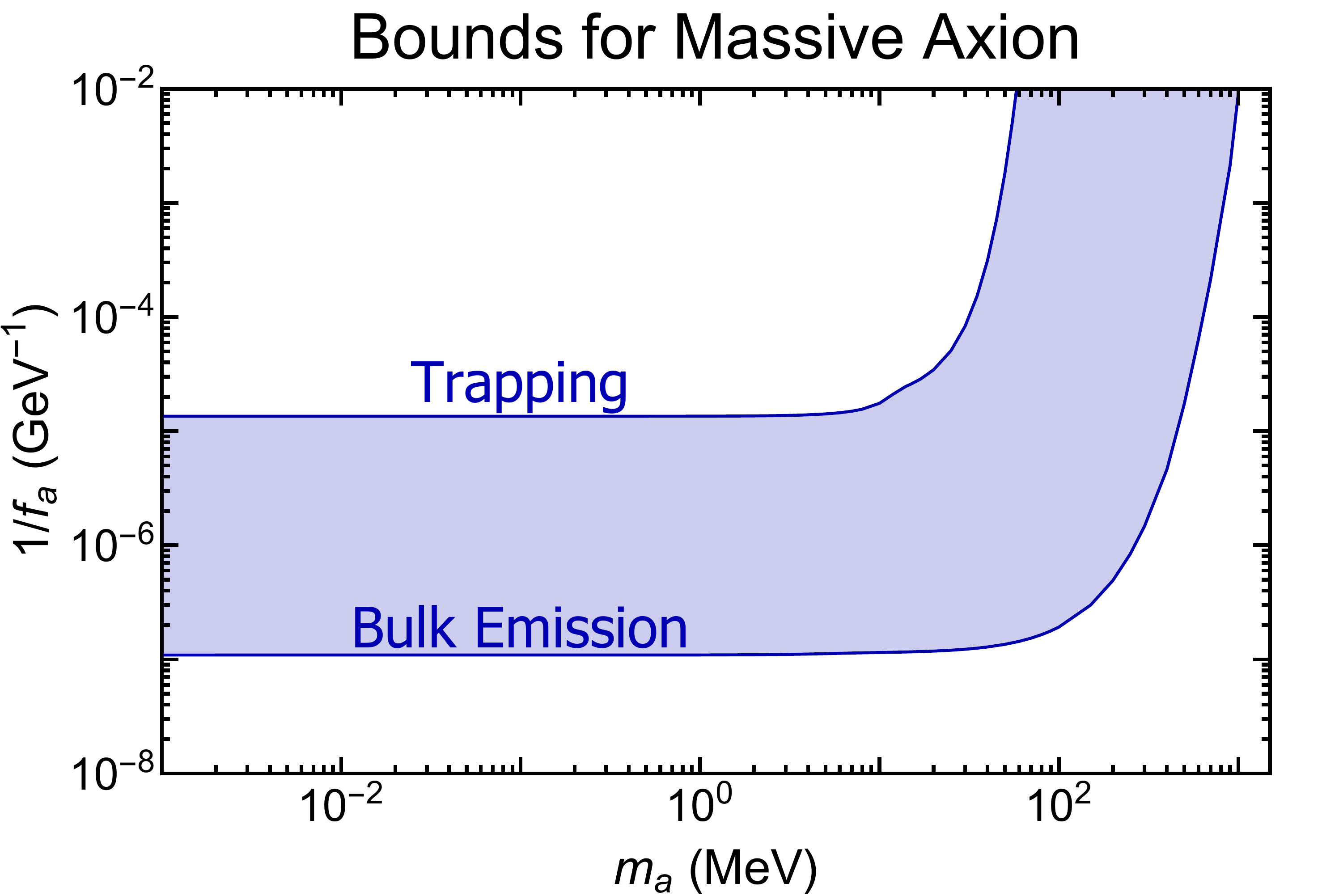}
\includegraphics[width=.49\linewidth]{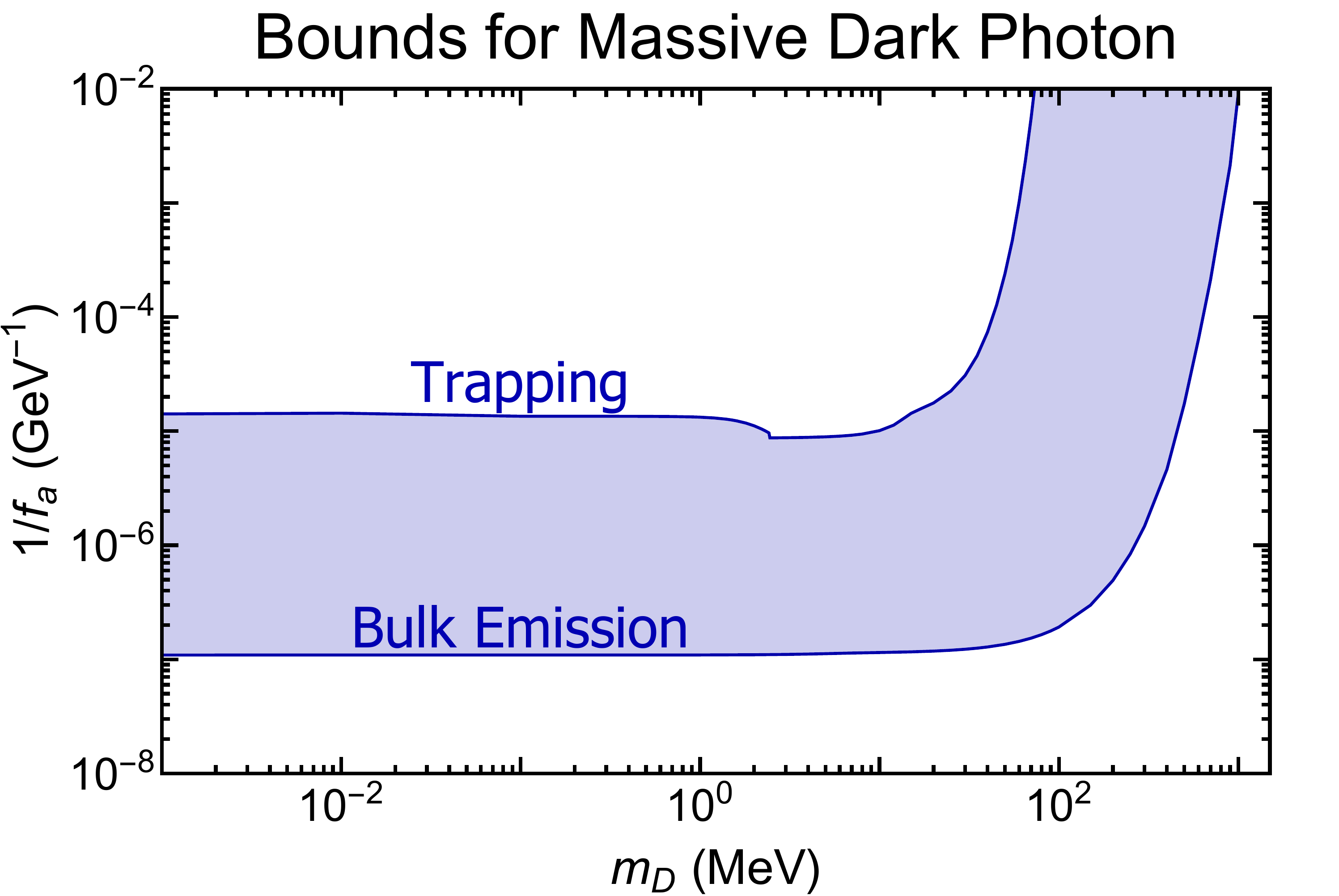}
\caption{These plots show the lower constraint from bulk emission and the upper constraint from trapping as a function of particle mass. The first plot shows these constraints for a massive axion and massless dark photon and the second plot shows them for a massive dark photon and a massless axion. Note that the upper trapping limit only contains the contributions from trapped axions and transverse dark photons since we must treat the longitudinal dark photons separately (See Sec.~\ref{Sec: Longitudinal}). }
\label{Fig: Supernova Bounds}
\end{figure}

The region of parameter space excluded by supernova is shown in Fig.~\ref{Fig: Supernova Bounds}. The first plot considers a massive axion and massless dark photon while the second considers a massless axion and massive dark photon. The high mass behavior of these constraints is quite different from the usual supernova constraint behavior. Ordinarily, the trapping constraint curve comes down to intersect the bulk emission curve to form a closed constrained region. However, we see that in our constraints, the region does not close off, but is an exponentially increasing strip. We will now look into the limiting behavior of the constraints in the high mass limit, $M\gg T$, where $M$ is the mass of the massive final state boson to determine the origin of this unique behavior.

\paragraph{Origin of the high mass scaling: }

The high mass behavior of the bulk emission curve is straightforward to understand. The initial particles must have enough energy to produce this very massive particle. This leads to the luminosity being Boltzmann suppressed for $M\gg T$.
The constraint curve is found by demanding $L_\text{bulk}(f)=L_\nu$ so in order to compensate for the exponentially decreasing luminosity, the coupling must increase in exactly the same way. This leads to the exponentially increasing constraint curve for the bulk emission shown in Fig.~\ref{Fig: Supernova Bounds}. This type of scaling is expected in supernova constraints for any novel massive particle.

The trapping constraint curve is a bit more difficult to understand. To understand the scaling at high mass, first consider a more typical model that contains only one novel particle radiating like a blackbody.  For this type of model, the constraint decreases with mass $M$ for $M \gg T$ due to the fact that the radiation of a blackbody decreases as the particles mass increases. In order to keep $L_{trapping}(r,M)=L_\nu$ the radius of the blackbody must decrease to compensate for this mass suppression. This smaller radius then implies that the constrained coupling is weaker and gives a decreasing coupling constraint in the high mass limit.

Contrast this behavior with a model, like ours, that contains two novel particles. As one of the particles gains mass, $M$, it is still true that the blackbody emission of that particle gets suppressed. However, the blackbody emission of the massless particle does not get suppressed. So in the high mass limit, the total luminosity is only due to blackbody radiation of this massless particle. Demanding that $L_{tot}=L_\nu$ then fixes the radius of emission to a radius $r_0$, which is independent of the heavy mass $M$.
Now, the trapping constraint curve is found by demanding that $\tau(r_0)=2/3$. This optical depth is determined by the strength of the scattering of the massless particle into the massive particle (with mass $M$). This up-scattering process will be Boltzmann suppressed for $M\gg T$ and so to compensate, the coupling must increase in exactly the same way. This gives the exponentially increasing constraint curve that we observe in Fig.~\ref{Fig: Supernova Bounds}.

We will now investigate in more detail the source of these two exponentially increasing curves to determine the shape of their fall off.  In particular, we will show that the bulk emission constraint scales exponentially in $M/T$ while the trapping constraint scales exponentially in $M^2/T^2$.  To do this we can look at the scaling of each constraint in the limit $M\gg T$ ignoring $\OO(1)$ factors. For both curves, we found that this scaling is the inverse of the Boltzmann suppression of the initial particles in each interaction. Let us start with the bulk emission constraint. The dominant process here is the annihilation process. The electrons and positrons are largely relativistic so that we can ignore their mass $m_e$. A simple computation gives,
$$M^2 > s \sim E_{e^+}E_{e^-}$$
This is most easily satisfied if both energies are of order $M$ leading to a Boltzmann suppression $e^{-M/T}$. Thus the bulk emission constraint scales like,
$$1/f_{a,bulk} \sim e^{M/T} \quad \text{for} \quad M\gg T.$$
For the trapping constraint, the dominant scattering is the scattering off of electrons. In order for this scattering to be kinematically allowed, $s>(M+m_e)^2$. Again, we ignore $m_e$ to find,
$$M^2>s\sim E_{e^-}E_b$$
Where $E_{e^-}$ is the energy of the initial electron and $E_b$ is the energy of the initial, massless boson. Since $E_b$ is fixed to be the thermal average energy, $E_b \sim T$, we conclude that 
$$E_{e^-}\gtrsim M^2/T.$$
This leads to a Boltzmann suppression of the scattering that scales as $e^{-M^2/T^2}$.  The scaling of the trapping constraint curve is thus
$$1/f_{a,trap}\sim e^{M^2/T^2} \quad \text{for}\quad M\gg T.$$
Since the trapping curve scales as $e^{M^2/T^2}$ and the bulk emission curve scales as $e^{M/T}$ the two curves never meet and diverge from one another. Note that if $E_b$ is not fixed to be the averaged energy, then we are free to let both $E_b\sim M$ and $E_{e^-}\sim M$ so that our Boltzmann suppression will look like $ e^{-M/T}$. This will lead to a trapping constraint that scales as $e^{M/T}$ for $M\gg T$. This very different asymptotic scaling of the trapping constraint highlights how important it is to use the averaged initial energy rather than averaging over the initial energies of the boson.

\section{Longitudinal Dark Photons} 
\label{Sec: Longitudinal}

A massive vector boson like the dark photon has a longitudinal polarization which behaves quite differently from its transverse counterparts. In particular, as we will show, for small dark photon mass, the coupling of the longitudinal dark photon is highly suppressed. This difference in coupling motivates us to consider the transverse and longitudinal dark photons as two separate particles. This allows us to get two constrained regions by applying the constraints from Sec.~\ref{Sec: SN Emission} and Sec.~\ref{Sec: Trapping} to each polarization separately. 

\begin{figure}[H]
\centering
\includegraphics[width=.7\linewidth]{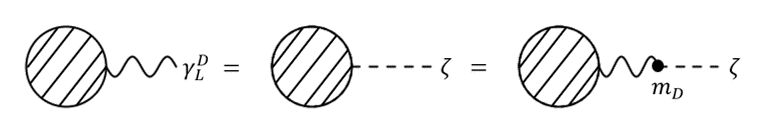}
\caption{A pictorial representation of the suppression of matrix elements containing longitudinal dark photons in the low mass/high energy limit. The first equality is the Goldstone boson equivalence theorem stating that in the low mass/high energy limit any matrix element with an external longitudinal photon is equal to the same matrix element with the longitudinal dark photon replaced with the Goldstone boson. The second equality makes use of the 2 point Goldstone bosons-dark photon interaction term that results from leaving unitarity gauge. This coupling scales as $m_D$ which gives the desired result.}
\label{Fig: Goldstone Equivalence}
\end{figure}

Let us now consider longitudinal dark photons in the small $m_D$ limit. We will show that the coupling of the longitudinal dark photons gets suppressed by a factor of its mass, $m_D$. This is a result of the Goldstone Boson Equivalence Theorem. To see this, consider leaving the unitary gauge of our model through a gauge transformation
\bea
A^D_\mu \rar {A}^D_\mu +\frac{1}{m_D}\partial_\mu \zeta \, .
\label{Eq: Leaving Unitarity}
\eea
The mass term for the dark photon gives rise to a kinetic term for the Goldstone boson, $\zeta$, and a 2-point interaction/mixing term with the dark photon
\bea
\frac{1}{2}m_D^2 A^\mu_DA^D_\mu \rar \frac{1}{2}m_D^2 A^\mu_DA^D_\mu+\frac{1}{2}\partial_\mu \zeta\partial^\mu\zeta +m_DA_D^\mu\partial_\mu \zeta .
\label{Eq: Nonunitarity Lagrangian}
\eea
The Goldstone Boson Equivalence Theorem states that in the low mass/high energy limit, the amplitude for a matrix element containing a longitudinal dark photon is equal to the amplitude for the same process with the longitudinal dark photon replaced with the Goldstone boson. In the low mass/high energy limit the mixing term $m_DA_D^\mu\partial_\mu \zeta$ can be treated as an interaction term with coupling $m_D$. Since this is the only interaction the Goldstone boson has, we know that any matrix element with an external Goldstone boson must scale as $m_D$. So for masses with $m_D\ll T$, we should find $\MM_L\sim m_D$ where $\MM_L$ is any matrix element involving a longitudinal dark photon in the initial or final state. This argument is illustrated pictorially in Fig.~\ref{Fig: Goldstone Equivalence}.

Now let us compute the constraints coming from the longitudinal dark photon. We start with the bulk emission constraint from Sec.~\ref{Sec: SN Emission}. In order to compute this bulk emission constraint we need to compute the emissivity of longitudinal dark photons
\bea
\dot \epsilon_L^D =\(\prod_i \int\frac{d^3 \vec p_i}{(2\pi)^3 2 E_i} f_i\)\(\prod_f \int\frac{d^3 \vec p_f}{(2\pi)^3 2 E_f}\)\avg{|\MM_L|^2}E_D(2\pi)^4\delta^4\(\sum p_i-\sum p_f\).
\label{Eq: Longitudinal Production}
\eea
This is exactly the same as Eq.~\ref{Eq: Emissivity General}, but rather than summing over all dark photons spins in the final state, we simply insert the longitudinal polarization for the dark photon (where longitudinal is defined in the frame of the star).  We also are only accounting for the energy emitted by the longitudinally polarized dark photons, since we will assume that the axion is trapped in the supernova for the couplings of interest. We will consider two process which produce longitudinal dark photons. First, we will consider the annihilation process, since it was the dominant process in the full emissivity. Second, we consider the plasmon decay process because the mass suppression has a smaller impact on this process compared to the others. A general computation given in App.~\ref{App: Computations} shows that any production process will have a matrix element that scales as $m_D^2/s$ in accordance with the Goldstone boson equivalence theorem.  For the plasmon decay, the center of mass energy in this process is the plasmon mass which is of the same order as the plasmon frequency, $\omega_p$. The plasmon frequency is around a full order of magnitude smaller than the typical center of mass energy for annihilation so the $m_D^2/s$ suppression factor leads to the plasmon decay emissivity being significantly less suppressed by the mass of the dark photon than the other processes. In fact after computation, we find that plasmon decay is the dominant process in the production of longitudinal dark photons if $m_D < \omega_p$. The emissivities are computed in App.~\ref{App: Computations}
\bea
\begin{aligned}
\dot \epsilon_a^L & =\(\frac{m_D^2}{\bra s \ket}\) \frac{\alpha_e} {24 f_a^2}n_{e^-}n_{e^+}\bra E_{tot}\ket Q_L^a(m_a,m_D) \, , \\
 \dot \epsilon_p^L & =\(\frac{m_D^2}{\omega_p^2}\)\frac{\zeta(3)T^3}{3\pi f_a^2}\(\frac{\omega_p^2}{4\pi}\)^2Q_L^p(m_a,m_D) \, .
\end{aligned}
\label{Eq: Longitudinal Emissivities}
\eea
Just as before $Q^a_L$ and $Q^p_L$ are $\OO(1)$ numbers for low $m_D$ and $m_a$ and are defined in App.~\ref{App: Computations}. Comparing Eq.~\ref{Eq: Longitudinal Emissivities} to the emissivities given in Eq.~\ref{Eq: Computed Emissivities 2 to N} and Eq.~\ref{Eq: Computed Emissivity Decay} for the annihilation and plasmon decay emissivities, we can see that they have the same form but with an additional $m_D^2/ s$-type suppression factor in both.

For the longitudinal trapping bounds we will need to compute the cross section for longitudinal dark photons scattering with fermions, as well as their decay width. The decay width for the longitudinal dark photon can be seen to be the same as the decay for the transverse dark photon by considering the decay in the dark photon's rest frame. Thus the decay width is the same as that given in Eq.~\ref{Eq: Decay Widths}.

Next we turn to the scattering cross section for longitudinal dark photons. The matrix element for this process is the same as the second diagram in Fig.~\ref{Fig: 2 to N Processes}, with the polarization of the initial dark photon chosen to be longitudinal. The computation is very similar to that of the full scattering cross section and is detailed in App.~\ref{App: Computations}. The result is
\bea
\sigma^{\gamma_D^L\rar a}_{\Delta \vec p}=\frac{\alpha_e}{3f_a^2}\(\frac{m_D^2}{s}\)\frac{s\beta'}{2E_AE_B|\vec v_A-\vec v_B|}\(\frac{2|\vec p|^2\sin^2(\theta)}{s\beta^2}\(I(m_D^2,m_a^2)-\frac{V\beta'}{3\beta} \)+\frac{V\beta'}{3\beta}\)  .
\label{Eq: Longitudinal Scattering}
\eea

\begin{figure}[t]
\centering
\includegraphics[width=.9\linewidth]{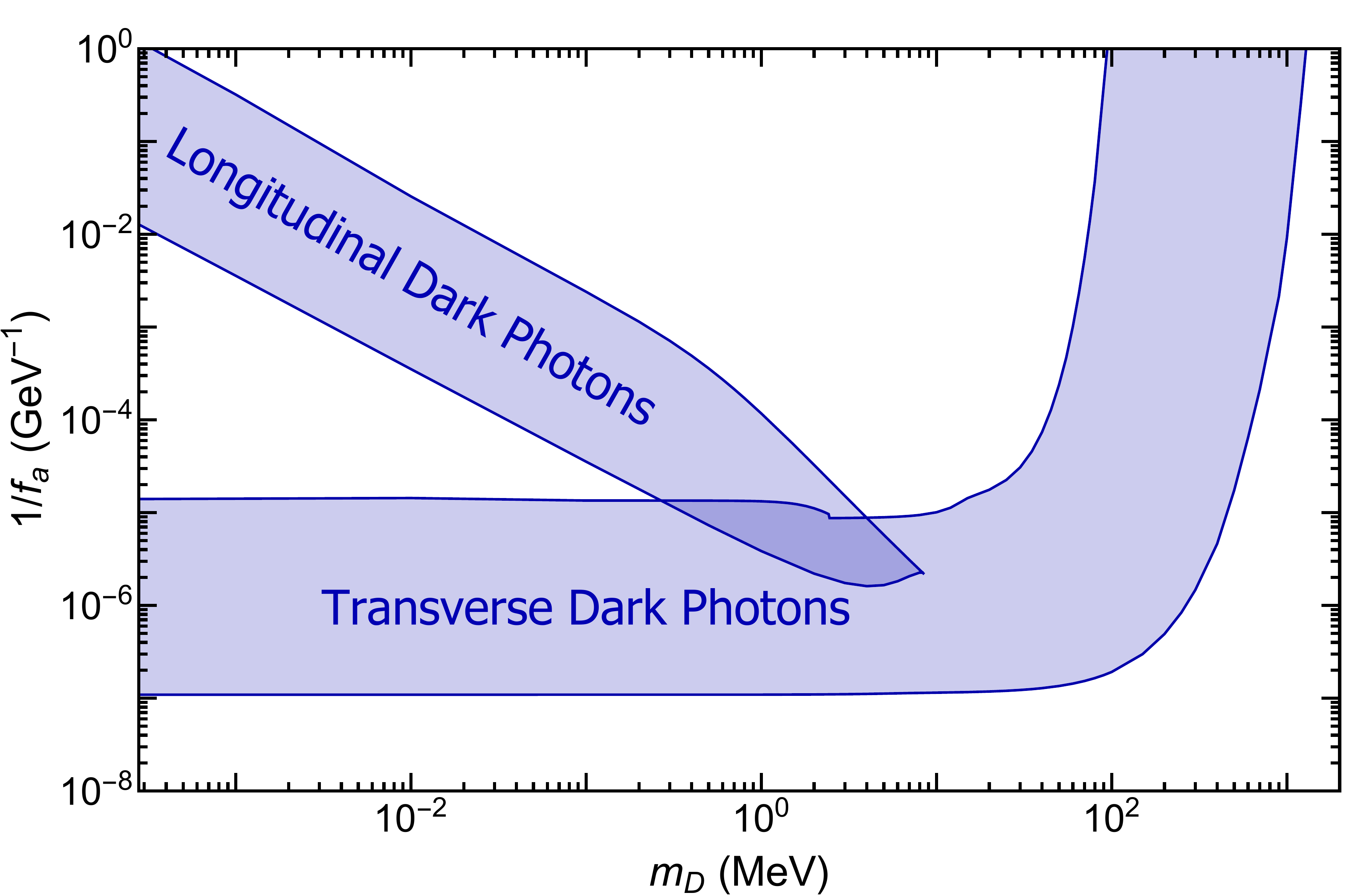}
\caption{A plot of the longitudinal and transverse constraints as a function of dark photon mass. The transverse constraints are analogous to those shown in Fig.~\ref{Fig: Supernova Bounds}. The longitudinal constraints are found by applying the bulk emission constraint and the trapping constraint to longitudinal dark photons only.}
\label{Fig: Longitudinal Constraints}
\end{figure}

Fig.~\ref{Fig: Longitudinal Constraints} shows the constraints for transverse dark photons and longitudinal dark photons described above as a function of the dark photon mass. The longitudinal constraints exhibit a number of interesting features. In the low mass regions, the longitudinal constraints on the coupling scale as $\sim m_D^{-1}$ due to the $\MM\sim m_D/f_a$ scaling described above. In the high mass region, this scaling changes for the longitudinal trapping as the decay becomes the dominant mode of trapping. The decay matrix element scales as $\MM \sim m_D^2$ and thus the scaling of the constraint on the coupling changes to $\sim m_D^{-2}$. The bulk longitudinal bound begins to increase at around $20$ MeV is due to the plasmon decay being kinematically forbidden once $m_D>m_{p}$, where $m_{p}\sim\omega_p$ is the mass of the plasmon.

\section{Other Astrophysical Constraints}
\label{Sec: WD and BBN}

As a final exercise, we can improve on our constraints, especially in the low mass regions by using our cross section computations in other astrophysical situations. We will examine white dwarf cooling bounds and BBN N$_{eff}$ bounds.

\paragraph{White Dwarfs} 
First we consider white dwarfs. A cooling bound on the coupling, similar to that found for SN1987A, can be derived from white dwarf data. For low temperature white dwarfs, the dominant source of energy emission is the emission of surface photons via blackbody radiation. This, along with the degeneracy of the white dwarf, allows one to derive a simple power law relationship between the luminosity of a white dwarf and the relative abundance of white dwarfs with that luminosity in the universe known as ``Mestel's Law".  For higher luminosity white dwarfs, bulk neutrino emission begins to play a significant role in cooling the white dwarfs resulting in a dip in the white dwarf abundance at higher luminosity. Observations of white dwarf abundance agree well with Mestel's Law and the so-called neutrino dip. An additional channel of significant bulk emission would increase this dip spoiling this agreement. It has been found that any additional mode of bulk particle emission must have an emissivity less than that of the neutrinos in order to preserve experimental agreement~\cite{Dreiner:2013tja}. Therefore, the cooling constraint for white dwarfs can be summarized as
\bea
\dot \epsilon_x<\dot \epsilon_\nu .
\label{Eq: White Dwarf Constraint}
\eea
For neutrino emission and our axion/dark photon emission, plasmon decay is the dominant source of particle production in white dwarfs and thus will be the only process considered.  Ref.~\cite{Raffelt:1996wa} has computed a nice expression for the emissivity of Standard Model neutrinos
\bea
\epsilon_\nu= \frac{8\zeta(3) T^3}{3\pi}\frac{C_V^2G_F^2}{\alpha_e}\(\frac{\omega_p^2}{4\pi}\)^3 Q_3,
\label{Eq: Neutrino Emissivity}
\eea
where 
$$Q_3=Q_3^L+Q_3^T$$
$$ Q_3^L=\frac{1}{4\zeta_3T^3}\int_0^{k_1}d|\vec k| \frac{|\vec k|^2}{e^{\omega_\ell/T}-1}\frac{\omega_\ell^2}{\omega_p^2}Z_L\hat \pi_\ell^{2} \, , \qquad Q_3^T=\frac{1}{2\zeta_3T^3}\int_0^{\infty}d|\vec k| \frac{|\vec k|^2}{e^{\omega_t/T}-1}Z_T\hat \pi_t^{3}.$$
This shares a similar form to our expression for the emissivity of axions and dark photons, given in Eq.~\ref{Eq: Computed Emissivity Decay}. In our constraints we take the white dwarf to have a constant temperature ($T=2$ keV) and constant plasmon frequency ($\omega_p =40$ keV) as is typical for  white dwarf constraints~\cite{Raffelt:1996wa}.  Applying Eq.~\ref{Eq: Computed Emissivity Decay} and Eq.~\ref{Eq: Neutrino Emissivity} to the constraint given in Eq.~\ref{Eq: White Dwarf Constraint} we get a constraint on the coupling
\bea
1/f_a<(2.3 \times 10^{-9}\;\text{GeV}^{-1})\sqrt{\frac{Q_3}{Q^p(m_a,m_D)}} \, ,
\label{Eq: White Dwarf Coupling}
\eea
in agreement with previous considerations of this bound~\cite{Pospelov:2018kdh,Arias:2020tzl}.
At low masses $Q_3\approx Q^p$, but as mass of the axion or dark photon increases $Q^p$ decreases.  An upper trapping bound on the excluded region from white dwarfs could be obtained but is rendered irrelevant by even the most conservative BBN $\Delta N_{eff}$ bound. 

\begin{figure}[t]
\centering
\includegraphics[width=.77\linewidth]{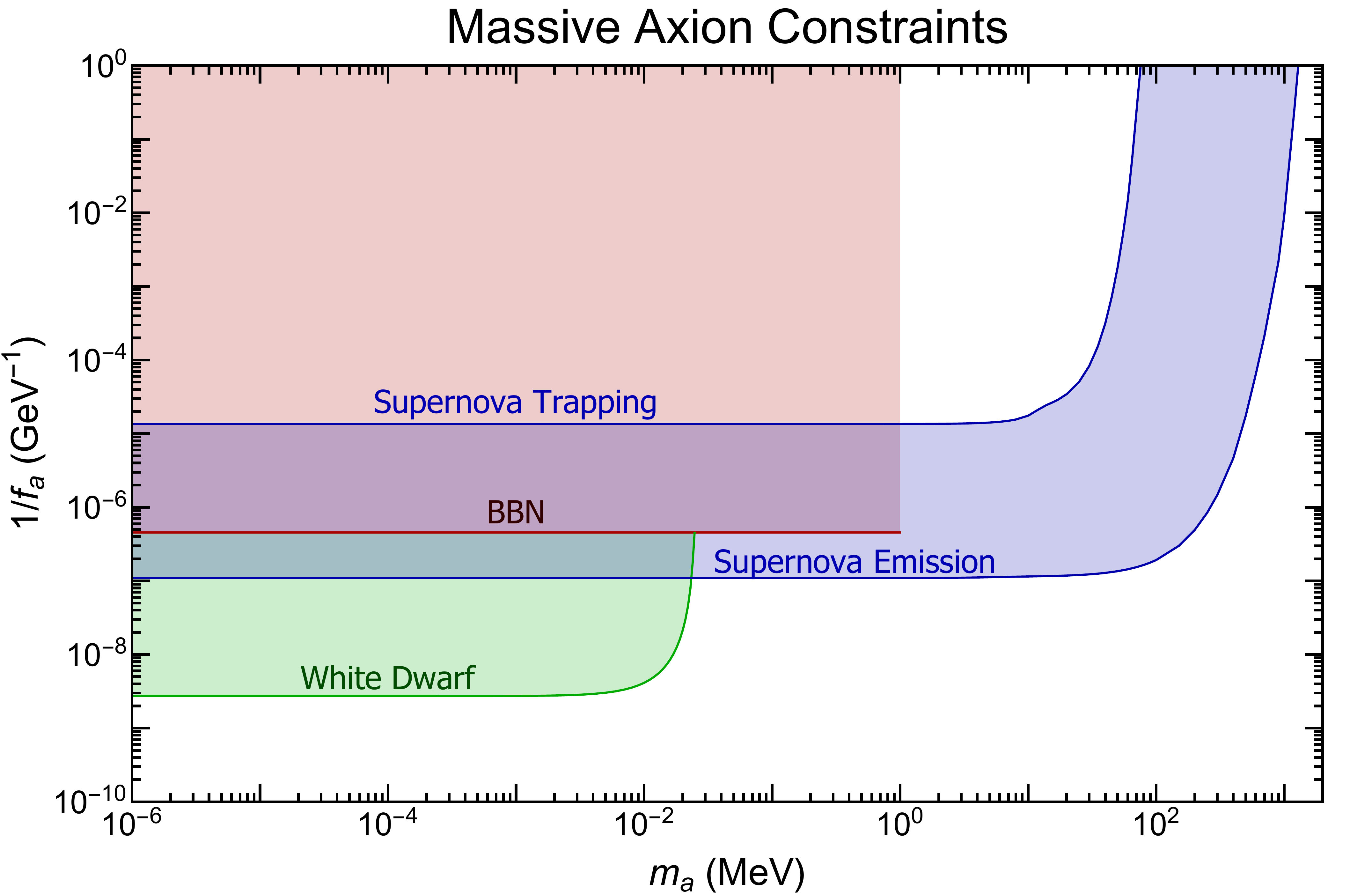}
\includegraphics[width=.77\linewidth]{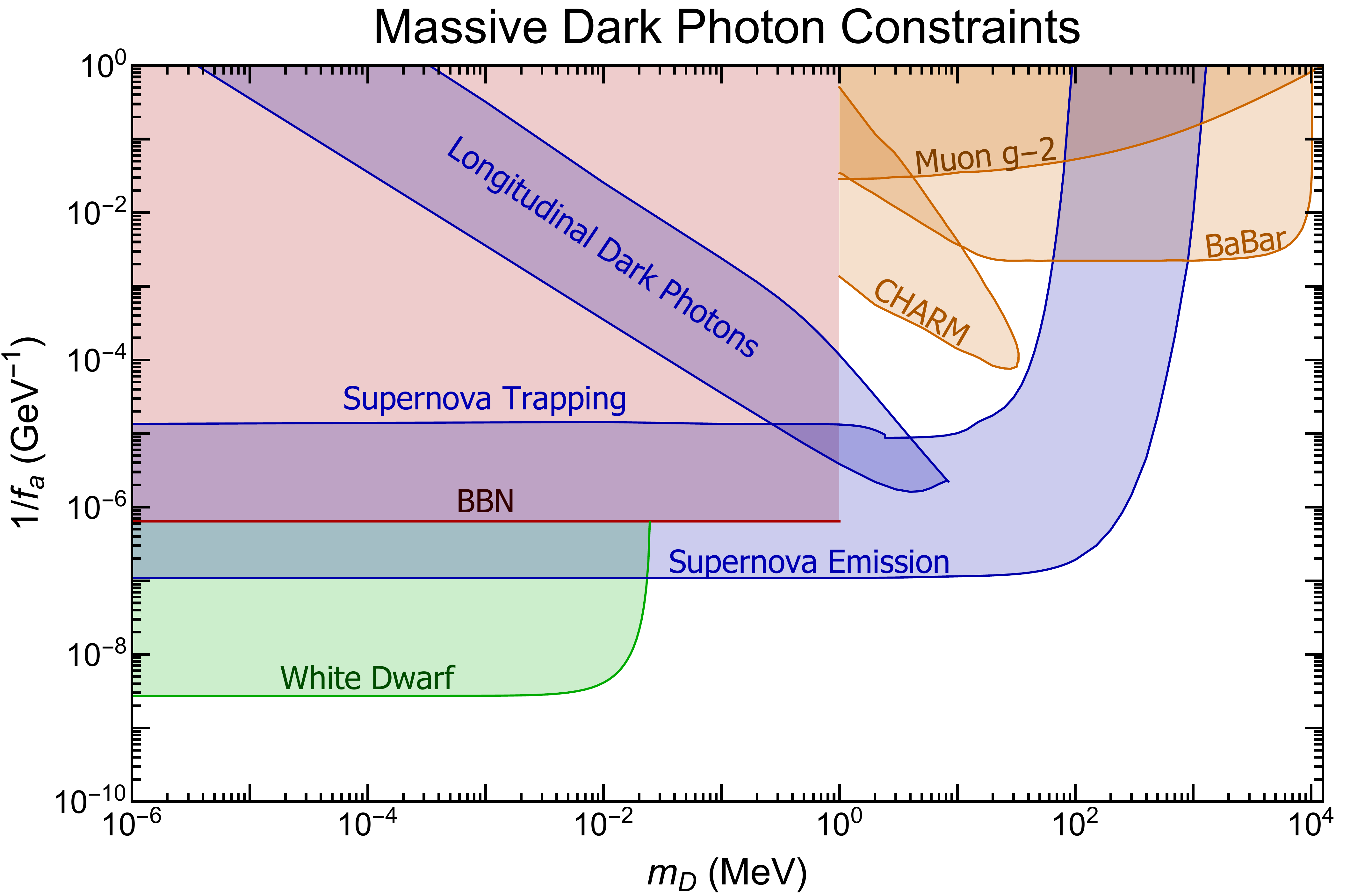}
\caption{The constrained regions on the coupling $1/f_a$ from the various astrophysical and collider sources. The first plot corresponds to a massive axion and a massless dark photon while the second plot corresponds to a massive dark photon and a massless axion. The blue constraint are those due to the supernova constraints examined in Sec.~\ref{Sec: SN Emission}, Sec.~\ref{Sec: Trapping} and Sec.~\ref{Sec: Longitudinal}. The green constraint is due to white dwarf cooling constraints examined in Sec.~\ref{Sec: WD and BBN}. The red constraint is a very conservative BBN constraint computed in Sec.~\ref{Sec: WD and BBN}. The orange constraints are the various collider constraints taken from Ref.~\cite{deNiverville:2018hrc}. Note that these constraints are only shown in the massive dark photon plot since  Ref.~\cite{deNiverville:2018hrc} only considers the $m_D\gg m_a$ limit. Similar collider constraints exist for the opposite limit but have not yet been computed. Also note that slightly stronger bounds than the white dwarf cooling bounds can be obtained by using the recent analysis of the tip of the red giant branch~\cite{Kalashev:2018bra,Capozzi:2020cbu}.}
\label{Fig: Full Constraints}
\end{figure}

\paragraph{BBN Bounds}
Now we consider a very simple, conservative BBN $\Delta N_{eff}$ bound. In our simple constraint we demand that axions and dark photons were not in equilibrium at nucleosynthesis as the addition of 3 (or 4 for massive dark photons) bosonic degrees freedom in the radiation density would induce too large of an increase in the number of effective neutrinos. We take the temperature at nucleosynthesis to be $T_{BBN} = 2$ MeV, both $m_a$ and $m_D$ to be smaller than $T_{BBN}$, and consider pair production of axions and dark photons through electron positron annihilation.
To implement $\Delta N_{eff}$ bounds, we demand that
\bea
n_e^2 \bra v_{mol} \sigma_{e^+e^-\rar a\gamma_D}\ket < n_{a,\gamma_D} H(T_{BBN}) ,
\label{Eq: NonEq Condition}
 \eea
where $n_e$ is the number density of electrons and $n_{a,\gamma_D}$ is meant to convey the equilibrium density of whichever particle is massless.
Neglecting the electron mass since $T_{BBN} > m_e$, we find
\bea
\bra  v_{mol} \sigma_{e^+e^-\rar a\gamma_D}\ket=\frac{\alpha_e}{24 f_a^2} .
\label{Eq: BBN cross section}
\eea
Inserting this with the densities and Hubble at 2 MeV, we get that the allowed couplings must satisfy,
$1/f_a< 4.5 \times 10^{-7} \; $GeV$^{-1}$
for massless axions and 
$1/f_a< 6.5 \times 10^{-7} \; $GeV$^{-1}$
for massless dark photons. This constraint applies until the mass of the axion or dark photon is around 1 MeV at which point they no longer contribute to the radiation density. 

These two constraints as well as the supernova and collider constraints taken from Ref.~\cite{deNiverville:2018hrc} are shown in Fig.~\ref{Fig: Full Constraints}. Note that only the constraints for the massive dark photon show collider constraints since  Ref.~\cite{deNiverville:2018hrc} derive their constraints explicitly in the limit $m_D\gg m_a$. Similar constraints could be derived in the limit $m_a\gg m_D$ although this has not been done. We expect that collider constraints for $m_a\gg m_D$ will exclude a similar region of the parameter space as those for $m_D\gg m_a$, but have refrained from including it in Fig.~\ref{Fig: Full Constraints}.

\section{Conclusion}
\label{Sec: conclusion}

In this article we considered the supernova bounds on the well motivated topological axion-photon-dark photon coupling.  This coupling is interesting because it doesn't require or generate kinetic mixing between the vectors, while at the same time it can be responsible for generating the abundance of dark matter. The supernova bound on this coupling highlighted two generic features that appear in supernova constraints.  

The first feature occurs when the interaction being probed at the supernova involves multiple new particles with different masses, an example familiar from similar to the dynamics in dark matter models involving co-annihilation~\cite{Griest:1990kh} or co-scattering~\cite{DAgnolo:2017dbv}. 
When this situation occurs, the production and scattering cross section decreases whenever one of the particles obtains a mass so that the bulk emission and trapping bounds never meet and instead both decrease exponentially.

The second feature is a bit more ubiquitous and involves the importance of treating
longitudinal and transverse polarizations of particles independently, with their own cooling and trapping
constraints.  Because the longitudinal mode coupling is $m_D/E$ suppressed in the low mass 
limit, it acts as its own weakly coupled particle that can cool supernova giving a cooling and trapping bound separate from the transverse modes.

Because these properties are generic, they will appear in all manners of motivated scenarios.
The most obvious model where a new feature occurs is in dark photon models with kinetic mixing.
If the dark photon is heavier than twice the electron mass, trapping is dictated by decay rather than scatter.  Since decay treats all polarizations the same, the differences between transverse and longitudinal modes doesn't become relevant until the dark photon mass is less than an MeV, where supernova bounds are superseded by other constraints.

\acknowledgments{
We thank Jae Hyeok Chang for useful discussion and comments on the draft.  AH, GMT and CR were supported in part by the NSF grants PHY-1914480, PHY-1914731, and by the Maryland Center for Fundamental Physics (MCFP). GMT is also partly supported by the US-Israeli BSF Grant 2018236.
}

\pagebreak 
\appendix

\section{Computational Details}
\label{App: Computations}

In this appendix, we provide details on how to calculate the averaged cross-sections used in the constraints. First, we define some notation to simplify the calculations. We will be frequently integrating over final states, for which we will use the shortened notation
\bea
 \int d\Pi_f(p_1,p_2,...p_n) \equiv \int \prod_{j=1}^n \frac{d^3\vec p_j}{(2\pi)^3 2E_j}(2\pi)^4\delta^4(q-\sum_{j=1}^n p_j).
\label{Eq: Final State Integral}
\eea
We use the bracket notation to refer to a thermal average over initial states $A$ and $B$. For some general function $g(\vec p_A, \vec p_B)$ of these momenta,  this average is defined as
\bea
\bra g(\vec p_A,\vec p_B)\ket \equiv \int \frac{d^3\vec p_Ad^3 \vec p_B f_A(E_A,T,\mu_A) f_B(E_B,T,\mu_B)}{(2\pi)^6n_An_B} g(\vec p_A,\vec p_B) ,
\label{Eq: Thermal Average}
\eea
where
$$
n_{A,B}=\int \frac{d^3\vec p_{A,B}f_{A,B}(E_{A,B},T,\mu_{A,B})}{(2\pi)^3} \qquad \qquad f(E,T,\mu)=\frac{g}{\text{Exp}((E-\mu)/T)\mp 1} .
$$
In order to capture kinematic features of most of our processes we define two functions $\alpha$ and $\beta$ as follows,
\bea
\alpha(s,m_1,m_2)\equiv 1+\frac{m_1^2-m_2^2}{s} \qquad 
\beta(s,m_1,m_2)\equiv\sqrt{1-\frac{2(m_1^2+m_2^2)}{s}+\frac{(m_1^2-m_2^2)^2}{s^2}}.
\label{Eq: alpha and beta}
\eea
Physically, in the center of mass frame of two particles with center of mass energy $\sqrt{s}$, the energy and momentum of the first particle are $E_1=\frac{\sqrt{s}}{2}\alpha(s,m_1,m_2)$ and  $|\vec p|=\frac{\sqrt{s}}{2}\beta(s,m_1,m_2)$, with similar expressions for the second particle.

\subsection{Dark Photon and Axion Production Emissivities}

In this section we give the details of the three emissivities in Eq.~\ref{Eq: Computed Emissivities 2 to N} and the emissivity in Eq.~\ref{Eq: Computed Emissivity Decay}.
We start with Eq.~\ref{Eq: Emissivity General}. As a convention, in all processes, the initial momenta will be indexed by capital letters while the final momenta will be indexed with a number.  $p_1$ and $p_2$ represent the axion and dark photon momentum respectively. $k$ will be the momentum of the photon that produces the axion and dark photon and $q$ will be the total momentum. We will see that the matrix element for all processes can be factorized as
\bea
i\MM^X=i\MM^X_\mu\AA^\mu \where \AA^\mu=\frac{\epsilon^{\mu\nu\alpha\beta}}{f_a}k_\nu {p_2}_\alpha \epsilon^D_\beta.
\label{Eq: M Factorization}
\eea
The label $X$ on $\MM_\mu^X$ will label the process, taking values $A$ for annihilation, $B$ for bremsstrahlung, $C$ for Compton and $P$ for plasmon decay. The factor $\AA_\mu$ contains all of the matrix element's dependence on the axion and dark photon final states. Writing our matrix element in this way allows us to factor out our final state phase space integral into two and write  Eq.~\ref{Eq: Emissivity General} as two separate integrals,
$$
\int d\Pi_f(p_1,p_2,p_3,...)|\MM^X|^2(E_1+E_2)=$$
$$\(\int d\Pi_f(k,p_3,...)\frac{d(k^2)}{2\pi}k^0\MM^X_\mu{\MM^X}^*_\nu\)\(\int d\Pi_f(p_1,p_2)\AA^\mu{\AA^*}^\nu\).$$

We now have one factor that contains all of the process-dependent information contained in $\MM^X_\mu$ and a second piece that contains all of the dependence on the axion and dark photon momentum. This second piece can be computed immediately,
\bea
\int d\Pi_2(p_1,p_2) \AA^\mu {\AA^*}^\nu=\frac{k^4\beta^3(k^2,m_a,m_D)}{48\pi f_a^2}\(\frac{k^\mu k^\nu}{k^2}-\eta^{\mu\nu}\).
\label{Eq: AA Integral}
\eea
We know that for every process (except plasmon decay) $k^\mu\MM^X_\mu=0$. So the first term doesn't contribute and we get a simple expression for the emissivity of $2 \rightarrow N$ processes
\bea
\dot \epsilon=n_An_B \bra \frac{1}{2E_A2E_B}\int d\Pi_f(k,p_3,...)\frac{d(k^2)}{2\pi}k^0(-|{\MM^X}|^2)  \frac{k^4\beta^3(k^2,m_a,m_D)}{48\pi f_a^2}\ket .
\label{Eq: Appendix Emiss}
\eea
In each process, we must identify the correct portion of the matrix element $\MM^X_\mu$ and insert it into this equation.  

\paragraph{Annihilation}
Following the first diagram from Fig.~\ref{Fig: 2 to N Processes}, the matrix element for the annihilation process is
$$i\MM=\bar v(p_B)(ie\gamma^\mu)u(p_A)\(\frac{-ig_{\mu\nu}}{k^2}\)\(\frac{i\epsilon^{\alpha\nu\beta\sigma}}{f_a}k_\alpha (p_2)_\beta\epsilon_\sigma^*(p_2)\).$$
Comparing this with Eq.~\ref{Eq: M Factorization}, we can identify,
 $$\MM^A_\mu=\frac{-e}{q^2}\bar v(p_B)\gamma_\mu u(p_A) \, ,$$
from which we find
\bea
\dot \epsilon_a=\frac{\alpha_e}{24 f_a^2} n_{e^+}n_{e^-}\bra (E_A+E_B) \frac{s+2m_e^2}{2E_AE_B}\beta^3(s,m_a,m_D)\ket.
\eea
Finally we define the factor 
\bea 
Q^a(m_a,m_D,T,\mu_e)\equiv \bra \frac{(E_A+E_B)}{\bra E_A +E_B\ket} \frac{s+2m_e^2}{2E_AE_B}\beta^3(s,m_a,m_D)\ket
\label{Eq: Qa} 
\eea
so that we may simply write
$$\dot \epsilon_a=\frac{\alpha_e}{24 f_a^2} n_{e^+}n_{e^-}\bra E_{tot}\ket Q^a \, , $$
where $E_{tot} = E_A + E_B$. It is worth nothing that in the limit $m_D=m_a=m_e=0$, $Q^a=1$. So, given that electrons and positrons are dominantly relativistic in the supernova, $Q^a$ is $\OO(1)$ for small $m_a$ and $m_D$.
 
\paragraph{Nuclear Bremsstrahlung}
In the soft radiation approximation, there are two diagrams which contribute to the amplitude for bremsstrahlung production:

\begin{figure}[H]
\centering
\includegraphics[width=.75\linewidth]{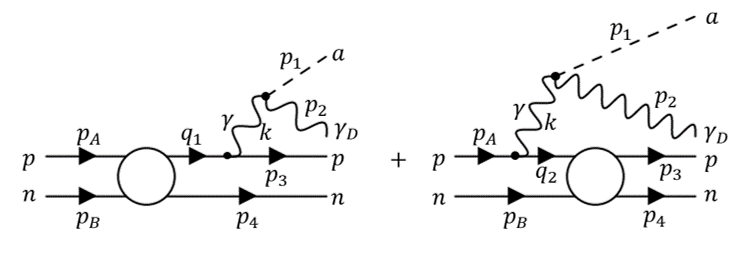}
\label{Fig: Brem Diagrams}
\end{figure}

In order to write down the matrix element for this process, we define a Dirac matrix $\Gamma_{pn}$ by
\begin{figure}[H]
\centering
\includegraphics[width=.6\linewidth]{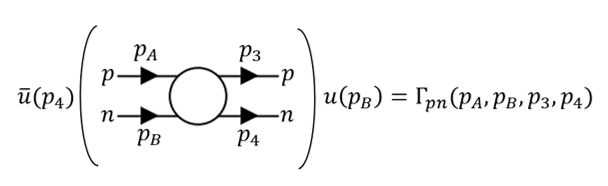}
\end{figure}
\noindent so that the amplitude for proton-neutron scattering can be written as
$$i\MM_{pn\rar pn}=\avg u(p_3) \Gamma_{pn}(p_A,p_B,p_3,p_4)u(p_A).$$
We can then write down the matrix element for our bremsstrahlung process
$$ i\MM=-\frac{ie}{k^2}\avg u(p_3)J^\mu u(p_A) \AA_\mu(k,p_2) ,$$
where
$$J^\mu=\gamma^\mu\(\frac{\slashed q_1+m_N}{q_1^2-m_N^2}\)\Gamma_{pn}(p_A,p_3)+\Gamma_{pn}(p_A,p_3)\(\frac{\slashed q_2+m_N}{q_2^2-m_N^2}\)\gamma^\mu.$$
So, we can identify the portion of the matrix element $\MM_B^\mu$ as
\bea
\MM_B^\mu=-\frac{e}{k^2}\avg u(p_3)J^\mu u(p_A) .
\eea
Because we are working in the soft radiation approximation we take $|\vec k| \ll |\vec p|$, where $\vec p$ is any of the nucleon momenta in the process. With that we get a simplified expression for $\MM_B^\mu$,
\bea
\MM_B^\mu=-e\MM_{pn}j^\mu \where j^\mu=2\(\frac{p_3^\mu}{k^2+2 k\cdot p_3}+\frac{p_A^\mu}{k^2-2k\cdot p_A}\)
\eea
We can then easily square this and sum over the remaining final spins and average over the initial spins. Using the fact that the nuclei are non-relativistic and expanding in their velocities we find
\bea
\avg{|\MM_B|^2}=-\frac{e^2}{k^4}\avg{|\MM_{pn}|^2}j^\mu j_\mu \where
j^\mu j_\mu=\frac{|\vec p_A-\vec p_3|^2}{(m_Nk^0)^2}\(1-\(\frac{|\vec k|}{k^0}\)^2\cos^2(\theta)\)
\eea
Next, we insert this into Eq.~\ref{Eq: Appendix Emiss},

\begin{equation}
\begin{aligned}
\dot \epsilon_B= &\;n_n n_p \bigg\langle \frac{1}{2E_A2E_B}\int \frac{d(k^2)}{2\pi}d\Pi_3(p_3,p_4,p_k)k_0 e^2\avg{|\MM_{pn}|^2}\\
 & \times \frac{|\vec p_A-\vec p_3|^2}{(m_Nk^0)^2}\(1-\(\frac{|\vec k|}{k^0}\)^2\cos^2(\theta)\)\(\frac{\beta^3(k^2,m_a,m_D)}{48\pi f_a^2}\)\bigg\rangle
\end{aligned}
\end{equation}
The remaining phase space integral will contain a delta function $\delta^4(p_A+p_B-p_3-p_4-k)$. Once again, we use the soft photon approximation to simplify and we ignore the photon momentum $k$ in the delta function. However, we must ensure that the photon energy is less than the center of mass energy of the nuclei, and so we impose, 
$$\delta^4(p_A+p_B-p_3-p_4-k)\rar \delta^4(p_A+p_B-p_3-p_4)e^{-k^0/T}.$$ 
This condition could have been achieved with a hard cutoff on the photon energy~\cite{Rrapaj:2015wgs}, however the exponential simplifies the computation and leads to similar results (see e.g.~\cite{Chang:2018rso,DeRocco:2019jti}). This substitution also allows us to factor the emissivity into two separate factors,
\begin{equation}
\begin{aligned}
\dot \epsilon_B= &\; n_n n_p\bra\frac{1}{2E_A2E_B}\int\Pi_3(p_3,p_4)\frac{|\vec p_A-\vec p_3|^2}{m_N^2}\avg{|\MM_{pn}|^2}\ket \\
 & \times \(\int \frac{d(k^2)d^3k}{(2\pi)^4 2k_0} k_0 \frac{e^2e^{-k^0/T}}{(k^0)^2}\(1-\(\frac{|\vec k|}{k^0}\)^2\cos^2(\theta)\)\(\frac{\beta^3(k^2,m_a,m_D)}{48\pi f_a^2}\)\).
\end{aligned}
\end{equation}
The first factor can be seen to be proportional to the momentum transfer cross section for $pn\rar pn$ scattering,
\bea
\frac{1}{2E_A2E_B}\int\Pi_3(p_3,p_4)\frac{|\vec p_A-\vec p_3|^2}{m_N^2}\avg{|\MM_{pn}|^2}=v_{mol}\frac{|\vec p_A|^2}{m_N^2}\sigma_{pn}^t\equiv\Sigma_{pn} \, ,
\label{Eq: Factored Brem Emissivity}
\eea 
where,
$$\sigma_{pn}^t=\int d\Omega (1-\cos(\theta))\frac{d\sigma_{pn}}{d\Omega} \, .$$

The second factor in Eq.~\ref{Eq: Factored Brem Emissivity} can be simplified significantly by performing the angular integrals and changing integration variables from $k^2$ and $\vec k$ to $k^0$ and $v=|\vec k|/k^0$. The result is, 
$$\int \frac{d(k^2)d^3k}{(2\pi)^4 2k_0} k_0 \frac{e^2e^{-k^0/T}}{(k^0)^2}\(1-\(\frac{|\vec k|}{k^0}\)^2\cos^2(\theta)\)\(\frac{\beta^3(k^2,m_a,m_D)}{48\pi f_a^2}\)$$
$$=\frac{\alpha_e}{12 f_a^2}\int \frac{dk^0dv}{4\pi^3}{k^0}^2e^{-k^0/T}v^2\(1-\frac{v^2}{3}\)\beta^3(k_0^2(1-v^2),m_a,m_D).$$ 
This integral can be computed exactly in the massless dark photon and axion limit,
\bea
\frac{\alpha_e}{12 f_a^2}\int \frac{dk^0dv}{4\pi^3}{k^0}^2e^{-k^0/T}v^2\(1-\frac{v^2}{3}\)=\frac{\alpha_eT^3}{90\pi^3 f_a^2}.
\label{Eq: Qb}
\eea
With this in mind we can write this second factor as $\frac{\alpha_eT^3}{90\pi^3 f_a^2}Q^b$, where 
$$Q^b(m_a,m_D,T)=\frac{15}{8T^3}\int dk_0dv e^{-k_0/T}k_0^2v^2\(1-\frac{v^2}{3}\)\beta^3(k_0^2(1-v^2),m_a,m_D).$$ 
Note that once again, $Q^b=1$ for $m_a=m_D=0$. The final expression for the emissivity is
\bea \dot \epsilon_b=\frac{\alpha_eT^3}{90\pi^3 f_a^2} n_p n_n \bra \Sigma_{pn}\ket Q^b(m_a,m_D,T).\eea

\paragraph{Compton}
There are two relevant diagrams for the Compton scattering process, 
\begin{figure}[H]
\centering
\includegraphics[width=.85\linewidth]{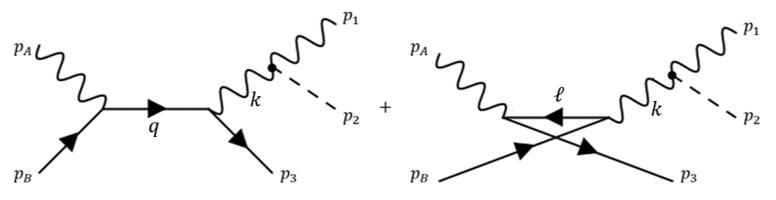}
\end{figure}
The matrix element is
$$i\MM=\frac{-ie^2}{k^2}\bar u(p_3)\(\frac{\gamma_\mu(\slashed q+m_e)\gamma^\nu}{q^2-m_e^2}+\frac{\gamma^\nu(\slashed \ell+m_e)\gamma_\mu}{\ell^2-m_e^2}\)u(p_B)\epsilon_\nu(p_A)\AA_\nu(k,p_2).$$
So we can identify the piece $\MM^C_\mu$ as
\begin{equation}
\MM^C_\mu=\frac{-ie^2}{k^2}\bar u(p_3)\(\frac{\gamma_\mu(\slashed q+m_e)\gamma^\nu}{q^2-m_e^2}+\frac{\gamma^\nu(\slashed \ell+m_e)\gamma_\mu}{\ell^2-m_e^2}\)u(p_B)\epsilon_\nu(p_A).
\end{equation}

Using this matrix element, and after some algebra we find 
\begin{equation}
\dot \epsilon_C=\frac{\alpha_e^2}{6f_a^2}n_{e^-}n_\gamma\bra \frac{1}{2E_AE_B}\int d(k^2)\beta^3(k^2,m_a,m_D)\int d\Pi_2(p_3,k)k^0 E(s,t,m_e,k^2)\ket \,
\end{equation}
where $t = l^2$, and
\begin{equation*}
\begin{aligned}
E(s,t,m_e,k^2)= & \left((2m_e^2+k^2)\(m_e^2\left(\frac{1}{s-m_e^2}+\frac{1}{t-m_e^2}\)^2+\(\frac{1}{s-m_e^2}+\frac{1}{t-m_e^2}\) \right. \right.  \\
& \left. \left. -\frac{k^2}{(s-m_e^2)(t-m_e^2)}\right) -\frac{1}{2}\(\frac{t-m_e^2}{s-m_e^2}+\frac{s-m_e^2}{t-m_e^2}\)\right).
\end{aligned}
\end{equation*}
After doing the integral over final states $d\Pi_2$, the end result is
\begin{equation}
\begin{aligned}
\dot \epsilon_C =  \frac{\alpha_e^2}{24\pi f_a^2}n_{e^-}n_\gamma & \bra \frac{1}{2E_AE_B}\int d(k^2)\beta^3(k^2,m_a,m_D)\left(\(\frac{s+m_e^2}{s-m_e^2}E_A-E_B\)R_1 \right. \right. \\ 
 & \left. \left. + \(\frac{k^2-2m_e^2}{s-m_e^2} E_A+E_B \)R_0 \right) \ket \, ,
\end{aligned}
\end{equation}
where,
$$R_0=\bigg(\frac{2s(2m_e^2+k^2)}{(s-m_e^2)^2}\(\beta(s,m_e,\sqrt{k^2})-\alpha(s,m_e,\sqrt{k^2})\text{arctanh}\(\frac{\beta(s,m_e,\sqrt{k^2})}{\alpha(s,m_e,\sqrt{k^2})}\)\)$$
$$+\frac{\alpha(s,m_e,\sqrt{k^2})\beta(s,m_e,\sqrt{k^2})}{4}+\text{arctanh}\(\frac{\beta(s,m_e,\sqrt{k^2})}{\alpha(s,m_e,\sqrt{k^2})}\)\bigg)$$
$$R_1=\bigg(\frac{2s(2m_e^2+k^2)}{(s-m_e^2)^2}\(\frac{m_e^2}{s}\text{arctanh}\(\frac{\beta(s,m_e,\sqrt{k^2})}{\alpha(s,m_e,\sqrt{k^2})}\)-\frac{\alpha(s,m_e,\sqrt{k^2})\beta(s,m_e,\sqrt{k^2})}{4}\)$$
$$+\frac{\beta^3(s,m_e,\sqrt{k^2})}{6}+\frac{\beta(s,m_e,\sqrt{k^2})}{2}\(1+\frac{m_e^2}{s}\)\bigg) \, .$$
The integral over $k^2$ runs from $(m_a+m_D)^2$ to $(\sqrt{s}-m_e)^2$. Unlike the other two processes, the $m_a=m_D=0$ limit does not lead to a simple expression. However, we may define
\begin{equation}
\begin{aligned}
	Q^c  = & \bra  \frac{s}{2E_AE_B}\int \frac{d(k^2)}{s}\beta^3\(\(\frac{s+m_e^2}{s-m_e^2}\frac{2E_A}{\bra E_{tot}\ket }-\frac{2 E_B}{\bra E_{tot}\ket }\)R_1 \right. \right. \\ 
	& \left. \left. + \(\frac{k^2-2m_e^2}{s-m_e^2}\frac{2 E_A}{\bra E_{tot}\ket }+\frac{2E_B}{{\bra E_{tot}\ket }}\)R_0\)\ket
\end{aligned}
\label{Eq: Qc}
\end{equation}
so we get,
\begin{equation}
\dot \epsilon_C=\frac{\alpha_e^2}{48\pi f_a^2}n_{e^-}n_\gamma \bra E_{tot}\ket Q^c(m_a,m_D,T,\mu_e).
\end{equation}
$Q^c$ is still roughly $\OO(1)$ in the limit $m_a=m_D=0$ since $E_A\approx \bra E_{tot}\ket /2$.

\paragraph{Plasmon Decay}
From the diagram in Fig.~\ref{Fig: Decay Process}, we can easily write the matrix element for plasmon decay as:
$$i\MM=-i\epsilon_\mu^P(k)\AA^\mu$$
Where $\epsilon_\mu^P$ is the plasmon polarization. If we square this and sum/average over final/initial spins and integrate over final states using the results of Eq.~\ref{Eq: AA Integral}, we can get
$$\int d\Pi_2(p_1,p_2) |\MM|^2=\(\frac{k^4 \beta^3(k^2,m_a,m_D)}{48\pi f^2}\)
\(\sum\epsilon_\mu^P(k){\epsilon^*}_\nu^P(k)\)\(\frac{k^\mu k^\nu}{k^2}-\eta^{\mu\nu}\).$$
We will compute the matrix element of longitudinally polarized plasmons and transversely polarized plasmons separately since their dispersion relations and field normalizations are different. The polarizations and dispersion relations are taken from Ref.~\cite{Braaten:1993jw}. The polarizations are
$$\epsilon^L_\mu=\frac{\omega_\ell}{|\vec{k}|}\sqrt{Z_\ell(k)}(1,\vec 0)\quad \epsilon^T_\mu =\sqrt{Z_t(k)}(0,\vec \epsilon_T).$$
Where $\vec \epsilon_T\cdot \vec k=0$. The dispersion relations are found by numerically solving, 
$$\omega_t^2(k)=k^2+\Pi_t(\omega_t(k),k) \quad \omega_\ell^2(k)=\frac{\omega_\ell^2(k)}{k^2}\Pi_\ell(\omega_\ell(k),k)$$
Where,
\begin{eqnarray*}
\Pi_t(\omega,k)&=& \omega_p^2\frac{3}{2v_*^2}\(\frac{\omega^2}{k^2}-\frac{\omega^2-v_*^2k^2}{k^2}\frac{\omega}{2v_*k}\ln\(\frac{\omega+v_*k}{\omega-v_*k}\)\) \, , \\  
\Pi_\ell(\omega,k) &=& \frac{3\omega_p^2}{v_*^2}\(\frac{\omega}{2 v_*k}\ln\(\frac{\omega +v_*k}{\omega-v_*k}\)-1\).
\end{eqnarray*}
These analytic equations for the dispersion relations are derived by assuming that the velocity of the particles in the plasma are dominated by single velocity, $v_*$ given by ${v_*}^2=\omega_1^2/\omega_p^2$, where
\bea
\omega_p^2=\frac{4\alpha}{\pi}\int_0^\infty dp\frac{p^2}{E}\(1-\frac{v^2}{3}\)\(n_{e^-}(E)+\avg n_{e^+}(E)\)
\label{Eq: Plasmon Frequency}
\eea
 $$ \omega_1^2=\frac{4\alpha}{\pi}\int_0^\infty dp\frac{p^2}{E}\(\frac{5}{3}v^2-v^4\)\(n_{e^-}(E)+\avg n_{e^+}(E)\).$$
This also defines the plasma frequency $\omega_p$. Finally, the field renormalizations for the plasmon are given by
$$Z_t=\frac{2\omega_t^2(\omega_t^2-v_*^2k^2)}{3\omega_p^2\omega_t^2+(\omega_t^2+k^2)(\omega_t^2-v_*^2k^2)-2\omega_t^2(\omega_t^2-k^2)}$$
$$ Z_\ell=\frac{2(\omega_\ell^2-v_*^2k^2)}{3\omega_p^2-(\omega_\ell^2-v_*^2k^2)}.$$
Using all of this, our longitudinal and transverse matrix elements are
$$\int d\Pi_2(p_1,p_2) |\MM|^2_{L}=\frac{Z_\ell \omega_\ell^2}{k^2}\(\frac{k^4 \beta^3(k^2,m_a,m_D)}{48\pi f_a^2}\)$$
\begin{equation} 
\int d\Pi_2(p_1,p_2) |\MM|^2_{T}=Z_t\(\frac{k^4 \beta^3(k^2,m_a,m_D)}{48\pi f_a^2}\)
\end{equation}
Then for each polarization
$$\omega_{\ell,t}\Gamma_{L,T} = \frac{1}{2}\int d\Pi_2(p_1,p_2) |\MM|^2_{L,T}.$$
If we define
$$\hat\pi_{\ell,t}\equiv \frac{\omega_{\ell,t}^2-|\vec k|^2}{\omega_p^2},$$
We can write the emissivity using Eq.~\ref{Eq: Emissivities Decay} 
\bea
\epsilon_a=\frac{\zeta_3T^3}{3\pi f_a^2}\(\frac{\omega_P^2}{4\pi}\)^2Q^P
\label{Eq: Qp}
\eea
With $Q^P=Q^P_L+Q^P_T$ where,
$$Q^P_L = \frac{1}{4\zeta_3T^3}\int_0^{k_1}d|\vec k| \frac{|\vec k|^2}{e^{\omega_\ell/T}-1}\frac{\omega_\ell^2}{\omega_p^2}Z_L\hat \pi_\ell \beta^3(\omega_p^2\hat \pi_t,m_a,m_D)$$
$$Q^P_T = \frac{1}{2\zeta_3T^3}\int_0^{\infty}d|\vec k| \frac{|\vec k|^2}{e^{\omega_t/T}-1}Z_T\hat \pi_t^2\beta^3(\omega_p^2\hat \pi_t,m_a,m_D).$$

\subsection{Scattering Cross Section}

The matrix elements for each process are shown in Fig.~\ref{Fig: Scattering Processes} and their amplitudes are simple to write, using notation developed in the previous sections. 
$$i\MM_{\psi\gamma_D\rar \psi a}=\frac{iQ_\psi}{q^2}\bar u(p')\gamma_\mu u(p)\AA^\mu(q,k')$$
$$i\MM_{\psi a \rar \psi \gamma_D}=\frac{-iQ_\psi}{q^2}\bar u(p')\gamma_\mu u(p){\AA^*}^\mu(q,k')$$
$Q_\psi$ is the electric charge of the fermion, $\psi$. We can easily square this matrix element, sum over final spins and average over initial state spins:

\bea
\avg{|\MM|^2}=\frac{2e^2}{g_{\star,k} t^2}\(2p_\mu p_\nu+\frac{t}{2}\eta_{\mu\nu}\)\sum_{\gamma_D \; spins}\AA^\mu(q,k'){\AA^*}^\nu(q,k')
\label {Eq: Scattering Matrix Element}
\eea
$g_{\star, k}$ represents the degrees of freedom in the initial particle that is either the axion or dark photon and $t$ is the Mandelstam variable equal to the square of the transferred momentum.  We will also use $k$ to denote the initial particle four momentum and $k'$ to denote the final particles four momentum regardless of if that particle is an axion or dark photon. Framing the computation this way allows us to compute both matrix elements at once. As discussed in Sec.~\ref{Sec: Trapping}, we are interested in computing the momentum transfer cross section defined in Eq.~\ref{Eq: Momentum Transfer Cross Section}. For computational ease we consider a relativistically invariant generalization of the momentum transfer cross section by replacing the $1-\cos(\theta)$ with $2t/(sA)$.  Where
$$t=(k-k')^2=\frac{s}{2}(A+B\cos(\theta))$$
and
$$ A=2\frac{k^2+{k'}^2}{s}-\alpha(s,\sqrt{k^2},m_\psi)\alpha(s,\sqrt{{k'}^2},m_\psi) \quad B=\beta(s,\sqrt{k^2},m_\psi)\beta(s,\sqrt{k'^2},m_\psi). $$
It is easy to see that in the limit $m_\psi/s\ll1$ (valid for electrons and positrons which are relativistic in the supernova) and the limit $m_\psi^2\gg k^2,{k'}^2$ (valid for protons)
$$B\rar -A\implies t\rar\frac{sA}{2}(1-\cos(\theta)) \implies 2t/(sA)\rar 1-\cos(\theta).$$ 
This property allows us to compute the momentum transfer cross section without sacrificing relativistic invariance. 
$$\sigma_{\Delta \vec p} \approx \frac{1}{4E_AE_B\Delta \vec v}\int d\Pi_f |\MM|^2 \(\frac{2t}{sA}\)$$
We have checked that the final results are insensitive to this particular deviation from standard convention.

In order to compute the cross section will need to compute the final state integral
$$R_1=\int d\Pi_2(k',p')t |\MM|^2$$

This integral can be computed by expanding in terms of Lorentz covariant tensors involving the initial state momenta and metric. We can then compute the necessary coefficients in this expansion by picking specific components for the indices in the center of mass frame or contracting. When all is done, the result is,
\begin{equation}
R_1=-\frac{2e^2}{g_{\star,k}f_a^2}\frac{s\beta'}{32\pi}\(k^2{\beta'}^2+{k'}^2\beta^2+\frac{s\beta^2{\beta'}^2}{3}+s\beta\beta'\(\frac{1}{V}-\frac{1-V^2}{V^2}\text{arctanh}(V)\)\)
\end{equation}
Where $$V=-B/A \quad \beta=\beta(s,\sqrt{k^2},m_\psi)\quad \beta'=\beta(s,\sqrt{{k'}^2},m_\psi).$$ 
Finally, plugging this into the cross section, we find
$$\sigma_{\Delta \vec p}=\frac{\alpha_e}{3 g_{\star,k}f^2}\frac{s\beta'}{2E_AE_B|\vec v_A-\vec v_B|}I(s,m_\psi^2,k^2,{k'}^2)$$
Where
\bea
I(s,m_\psi^2,k^2,k'^2)=\frac{3}{4}\(V\(\frac{k^2\beta'}{s\beta}+\frac{{k'}^2\beta}{s\beta'}+\frac{\beta{\beta'}}{3}\)+\(1-(1-V^2)\frac{\text{arctanh}(V)}{V}\)\).
\label{Eq: I Defintion}
\eea
Inserting $m_a$ and $m_D$ into the appropriate spots for each cross section gives us Eq~\ref{Eq: Scattering Cross Sections}.
 
\subsection{Longitudinal Scattering Cross Section}
In order to compute the longitudinal bounds, we need to compute the cross section for $\gamma_D^L e^-\rar a e^-$. The method described above transfers nicely to this specific case. Starting with Eq.~\ref{Eq: Scattering Matrix Element}, we replace the sum over dark photon polarizations with just the longitudinal polarizations. Since the dark photon is in the initial state here, these polarization don't interfere with the final state integrals which are expanded and computed just as before. The only remaining difference is the contraction with the epsilon tensors which now must be contracted with the explicit longitudinal polarization rather than the metric. After some simplification, the result is
$$\sigma^{\gamma_D^L\rar a}_{\Delta \vec p}=\frac{\alpha_e}{3f_a^2}\(\frac{m_D^2}{s}\)\frac{s\beta'}{2E_AE_B|\vec v_A-\vec v_B|}\(\frac{2|\vec p|^2\sin^2(\theta)}{s\beta^2}\(I(s,m_\psi^2,m_D^2,m_a^2)-\frac{V\beta'}{3\beta}\)+\frac{V\beta'}{3\beta}\).$$

\subsection{Decay Widths}
Let us now compute Eq.~\ref{Eq: Decay Widths} From the diagram given in Fig.~\ref{Fig: Scattering Processes} it is easy to see that the amplitude for scattering can be written as
$$i\MM=i\epsilon_\gamma^\mu\AA_\mu(p_1,p_2).$$
Here, $p_1$ will be the photon momentum and $p_2$ will be the other final state particle, axion or dark photon. Squaring this, summing over final spins, averaging over initial spins, and integrating over final states is straight forward.
$$\int d\Pi_f \sum \avg{|\MM_{a\rar D}|^2}=\frac{m_a^4}{16\pi f_a^2}\(1-\frac{m_D^2}{m_a^2}\)^3$$
$$\int d\Pi_f \sum \avg{|\MM_{D\rar a}|^2}=\frac{m_D^4}{16\pi f_a^2}\(1-\frac{m_a^2}{m_D^2}\)^3.$$
To find the decay width we simply divide by $2E$ where $E$ is the intial energy and we recover Eq.~\ref{Eq: Decay Widths}.
$$\Gamma_a=\frac{m_a^4}{32 \pi f_a^2 E}\(1-\frac{m_D^2}{m_a^2}\)^3 \quad \quad \Gamma_D=\frac{m_D^4}{96\pi f_a^2 E}\(1-\frac{m_a^2}{m_D^2}\)^3$$

\subsection{Production of Longitudinal Dark Photons}
The emissivity for a given process of longitudinal dark photons is given by Eq.~\ref {Eq: Longitudinal Production}.
Just as before, we separate the matrix element as $$i\MM_L=\MM^X_\mu \AA^\mu_L \where \AA_L^\mu=\frac{\epsilon^{\mu\nu\alpha\beta}}{f_a}k_\nu {p_2}_\alpha {\epsilon^D_L}_\beta.$$
For the two processes we are considering here, the final state is only the axion and dark photon so we can write
$$\dot\epsilon^D_L=\(\prod_i n_i\)\bra \MM^X_\mu\MM^X_\nu \int d\Pi_2(p_1,p_2)\AA_L^\mu\AA_L^\nu p_2^0\ket $$ 
Where $\prod_i n_i$ is the product of the densities of the initial state(s). The final state integral can be computed by considering the Lorentz covariant tensor
$$L^{\mu\nu\alpha}\equiv \int d\Pi_2(p_1,p_2)\AA_L^\mu\AA_L^\nu p_2^\alpha$$
of which we wish to take the $\alpha=0$ component. This integral can be computed by contracting the $\mu$ and $\nu$ indices, computing in the supernova frame where the longitudinal polarization takes the simple form $\epsilon_L=\frac{1}{m_D}\(|\vec k|,\hat{ \vec k}k^0\)$, and then properly relating the supernova frame momenta to the center of mass frame momenta so that we may do the final state integral in the center of mass frame. The result is that,
$$L^{\mu\nu\alpha}=L\(\frac{k^\mu k^\nu}{k^2}-\eta^{\mu\nu}\)k^\alpha$$
Where,
$$L =\frac{\beta_D^3 k^4}{48 \pi f_a^2} k^0 \frac{v^2 m_D^2}{k^2}\(\alpha_D I_0 + v \beta_D I_1\)$$
And
$$I_n\equiv \int_{-1}^1\frac{x^n dx}{\frac{(\beta_D+ v \alpha_D x)^2}{1-x^2}+\frac{4 v^2 m_D^2}{k^2}} \quad \beta_D=\beta(k^2,m_D,m_a)$$ 
$$\alpha_D=\alpha(k^2,m_D,m_a) \quad v=\frac{|\vec k|}{k^0}.$$
The rest of the computations follow just as before. The end result is,
\begin{equation}
\dot \epsilon_a^L =\frac{m_D^2}{\bra s \ket} \frac{\alpha_e} {24 f^2}n_{e^-}n_{e^+}\bra E_{tot}\ket Q_L^a(m_a,m_D) \quad \dot\epsilon_p^L=\frac{\zeta(3)T^3}{3\pi f_a^2}\(\frac{\omega_p^2}{4\pi}\)^2\(\frac{m_D^2}{\omega_p^2}\)Q_L^P(m_a,m_D) 
\end{equation}
Where, 
$$ Q_L^a(m_a,m_D)= \bra \frac{\bra s \ket v^2}{ s}\(\alpha_D I_0 + v \beta_D I_1\) \frac{(E_A+E_B)}{\bra E_A +E_B\ket}\frac{s+2m_e^2}{2E_AE_B}\beta^3(s,m_a,m_D)\ket$$
$$Q_L^P(m_a,m_D)=Q_{L,T}^p(m_a,m_D)+Q_{L,L}^p(m_a,m_D)$$

$$Q^P_{L,L}\equiv\frac{1}{4\zeta_3T^3}\int_0^{k_1}d|\vec k| \frac{|\vec k|^2}{e^{\omega_\ell/T}-1}\frac{\omega_\ell^2}{\omega_p^2}\frac{Z_L}{\hat \pi_\ell}\frac{|\vec k|^2}{\omega_\ell^2} \beta^3(\omega_p^2\hat \pi_t,m_a,m_D)\(\alpha_D I_0 + v \beta_D I_1\)$$
\begin{equation}
Q^P_{L,T}\equiv\frac{1}{2\zeta_3T^3}\int_0^{\infty}d|\vec k| \frac{|\vec k|^2}{e^{\omega_t/T}-1}\frac{|\vec k|^2}{\omega_t^2} Z_T \beta^3(\omega_p^2\hat \pi_t,m_a,m_D)\(\alpha_D I_0 + v \beta_D I_1\).
\end{equation}

\section{Supernova Profiles}
\label{App: Profiles}

\begin{figure}[H]
\centering
\includegraphics[width=.32\linewidth]{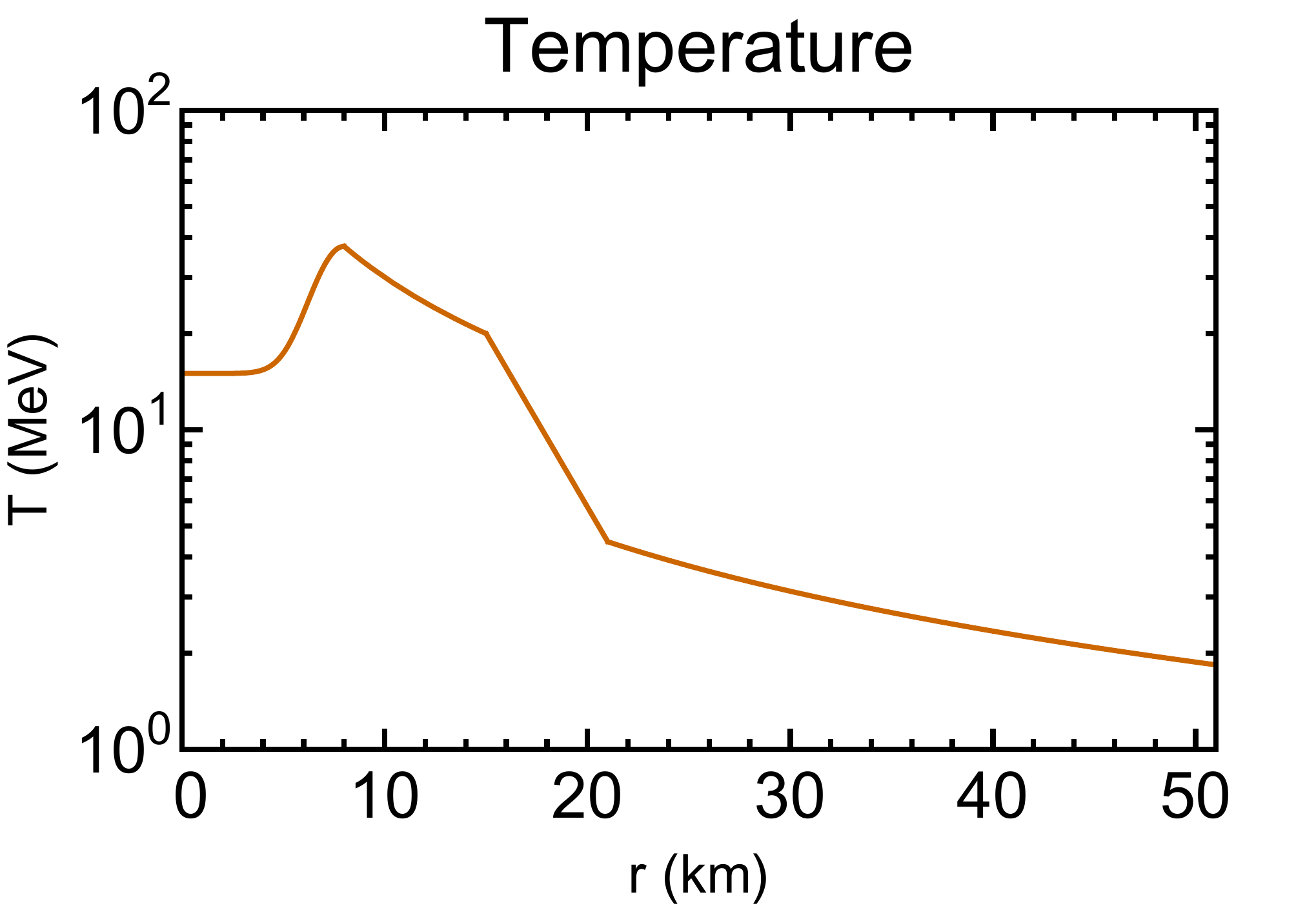}
\includegraphics[width=.32\linewidth]{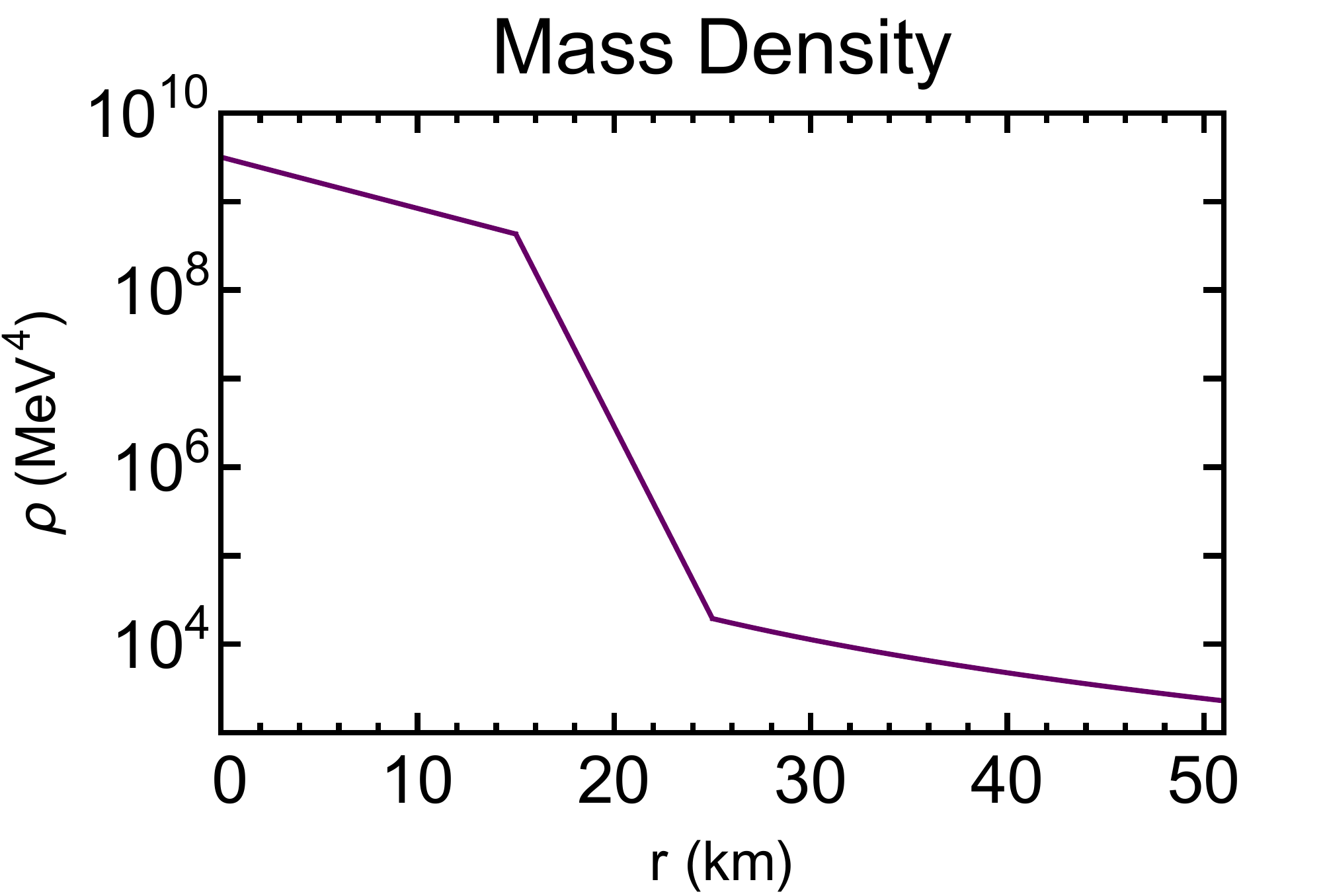}
\includegraphics[width=.32\linewidth]{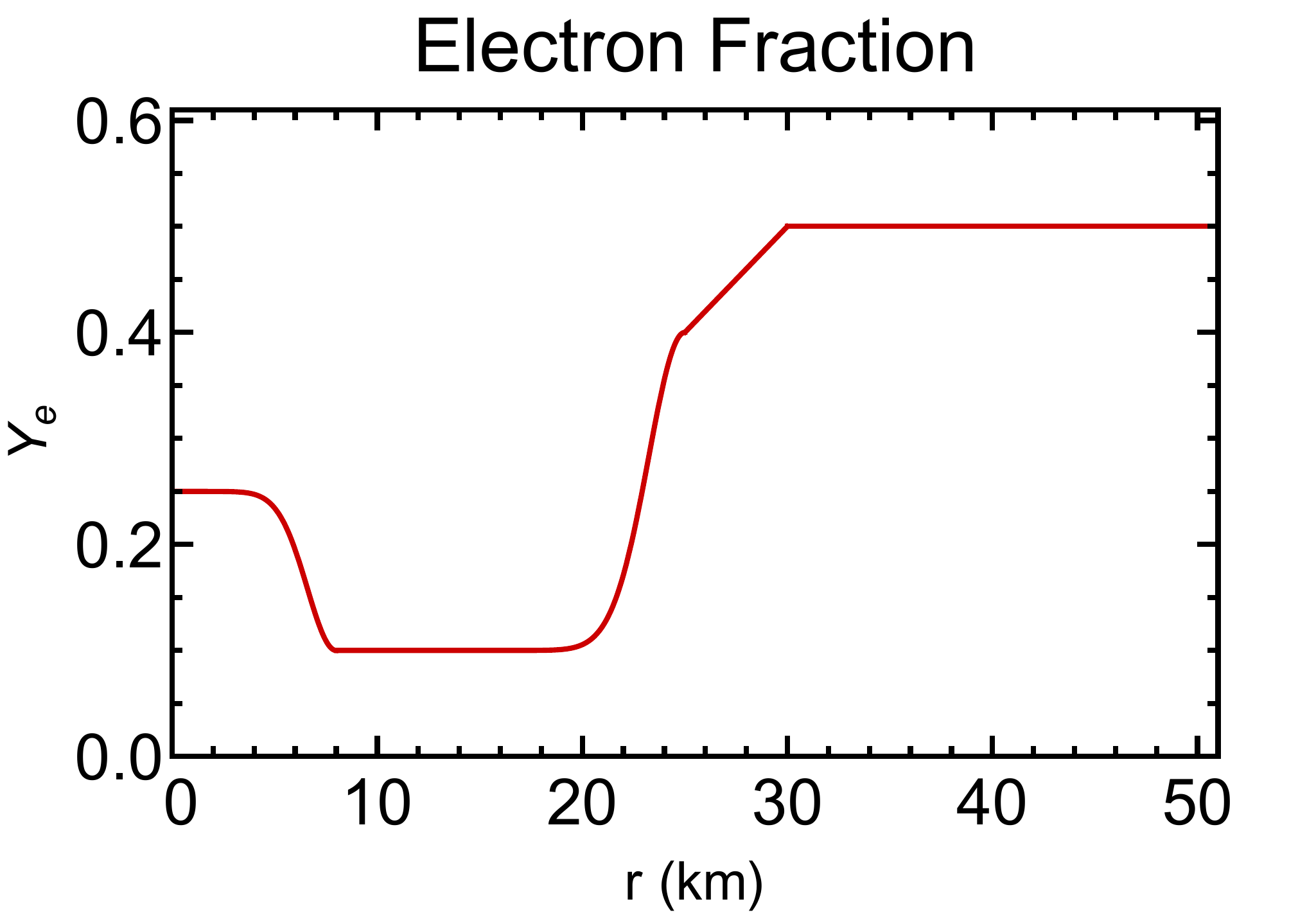}
\caption{The temperature, mass density, and electron fraction as a function of radius of the supernova.}
\label{Fig: Original Profiles}
\end{figure}

In this section we discuss the profile used in placing our constraints and the variation in the constraints placed that arise when using a different profile.  The temperature ($T$), mass density ($\rho$), and electron fraction ($Y_e$) profiles used in our computations use analytical fits to simulation used in Ref.~\cite{DeRocco:2019jti}.  These profiles are shown in Fig.~\ref{Fig: Original Profiles}. 
Our expressions for the emissivities and scattering cross sections involve number densities and chemical potentials.
From these profiles we can calculate the number densities for the electron, proton, neutron, photon and positron.  From the number densities, we can compute the chemical potentials of the electron, proton and neutron ($\mu_e$, $\mu_p$ and $\mu_n$ respectively).

\begin{figure}[H]
\centering
\includegraphics[width=.8\linewidth]{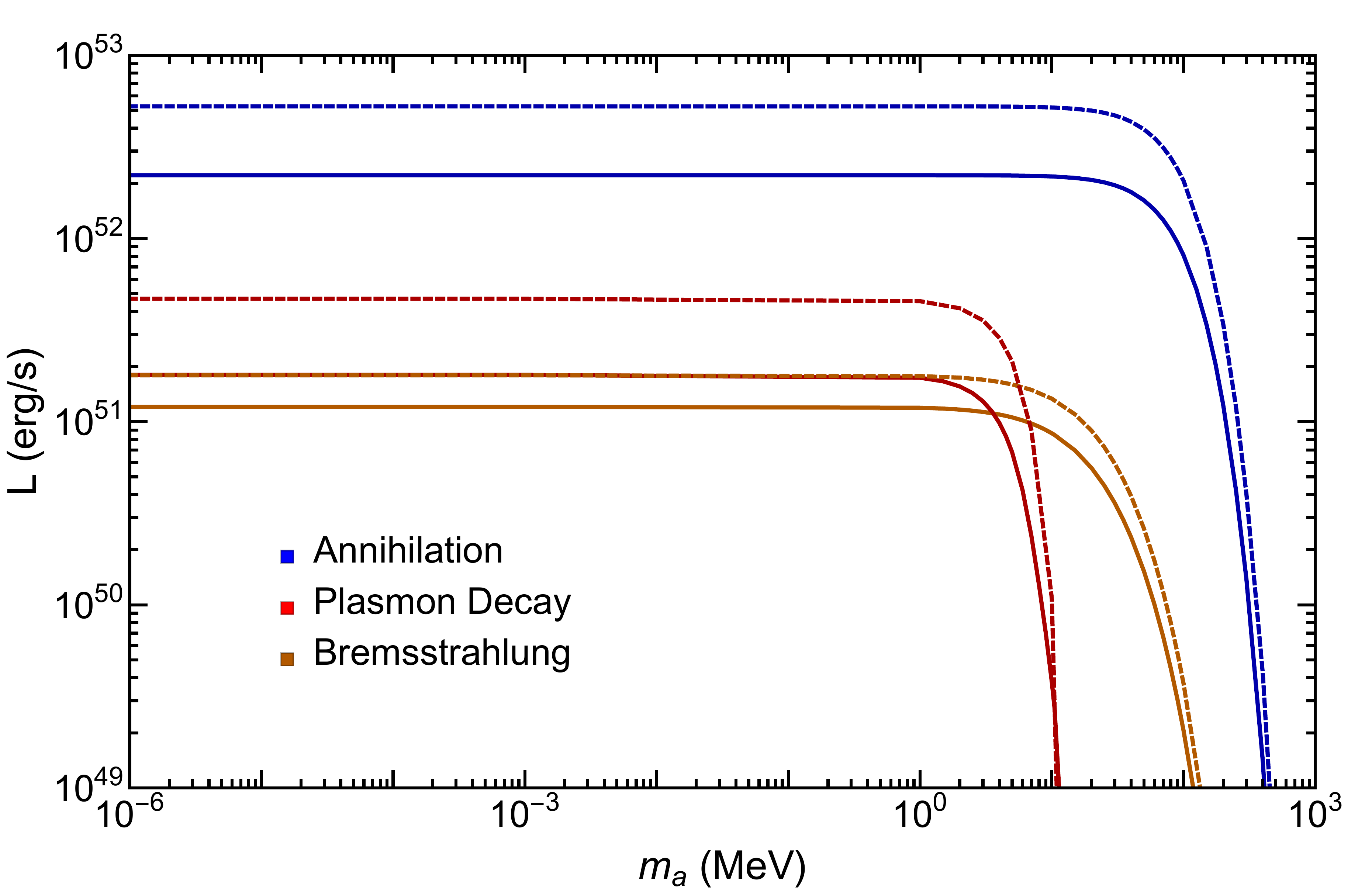}
\caption{The total luminosity of various channels as a function of axion mass.  The solid lines show the profile in Ref.~\cite{DeRocco:2019jti} while the dashed lines show the profile in Ref.~\cite{Fischer:2016cyd}.  The difference in the bounds on $f_a$ is about a factor of 1.5.}
\label{Fig: profiles}
\end{figure}

In order to access the dependence of our results on the profile used, we redid our calculations using the $18 \, M_\odot$ profile found in Ref.~\cite{Fischer:2016cyd}.
To compare the results visually, we show the total luminosity as a function of axion mass for the two different profiles in Fig.~\ref{Fig: profiles}.  In both profiles, annihilation is the dominant source of cooling.  The difference between the bulk cooling bound placed on $f_a$ when using the two different profiles is about a factor of 1.5, which is within the expected uncertainty of cooling bounds from supernovae.

\bibliographystyle{JHEP}
\bibliography{draft}{}

\providecommand{\href}[2]{#2}\begingroup\raggedright\begin{thebibliography}{10}

\bibitem{Hirata:1987hu}
{\bf Kamiokande-II} Collaboration, K.~Hirata et~al., {\it {Observation of a
  Neutrino Burst from the Supernova SN 1987a}},  {\em Phys. Rev. Lett.} {\bf
  58} (1987) 1490--1493.

\bibitem{Bionta:1987qt}
R.~M. Bionta et~al., {\it {Observation of a Neutrino Burst in Coincidence with
  Supernova SN 1987a in the Large Magellanic Cloud}},  {\em Phys. Rev. Lett.}
  {\bf 58} (1987) 1494.

\bibitem{Alekseev:1987ej}
E.~N. Alekseev, L.~N. Alekseeva, V.~I. Volchenko, and I.~V. Krivosheina, {\it
  {Possible Detection of a Neutrino Signal on 23 February 1987 at the Baksan
  Underground Scintillation Telescope of the Institute of Nuclear Research}},
  {\em JETP Lett.} {\bf 45} (1987) 589--592.

\bibitem{Burrows:1986me}
A.~Burrows and J.~M. Lattimer, {\it {The birth of neutron stars}},  {\em
  Astrophys. J.} {\bf 307} (1986) 178--196.

\bibitem{Burrows:1987zz}
A.~Burrows and J.~M. Lattimer, {\it {Neutrinos from SN 1987A}},  {\em
  Astrophys. J. Lett.} {\bf 318} (1987) L63--L68.

\bibitem{Raffelt:1996wa}
G.~G. Raffelt, {\em {Stars as laboratories for fundamental physics}: {The
  astrophysics of neutrinos, axions, and other weakly interacting particles}}.
\newblock 5, 1996.

\bibitem{Turner:1987by}
M.~S. Turner, {\it {Axions from SN 1987a}},  {\em Phys. Rev. Lett.} {\bf 60}
  (1988) 1797.

\bibitem{Raffelt:1987yt}
G.~Raffelt and D.~Seckel, {\it {Bounds on Exotic Particle Interactions from SN
  1987a}},  {\em Phys. Rev. Lett.} {\bf 60} (1988) 1793.

\bibitem{Raffelt:1988py}
G.~G. Raffelt, {\it {SUPERNOVA SN1987A AND SOME PROPERTIES OF LIGHT, EXOTIC
  PARTICLES}},  in {\em {24th International Conference on High-energy
  Physics}}, 8, 1988.

\bibitem{Lucente:2020whw}
G.~Lucente, P.~Carenza, T.~Fischer, M.~Giannotti, and A.~Mirizzi, {\it {Heavy
  axion-like particles and core-collapse supernovae: constraints and impact on
  the explosion mechanism}},  {\em JCAP} {\bf 12} (2020) 008,
  [\href{http://arxiv.org/abs/2008.04918}{{\tt arXiv:2008.04918}}].

\bibitem{Ishizuka:1989ts}
N.~Ishizuka and M.~Yoshimura, {\it {Axion and Dilaton Emissivity From Nascent
  Neutron Stars}},  {\em Prog. Theor. Phys.} {\bf 84} (1990) 233--250.

\bibitem{Bjorken:2009mm}
J.~D. Bjorken, R.~Essig, P.~Schuster, and N.~Toro, {\it {New Fixed-Target
  Experiments to Search for Dark Gauge Forces}},  {\em Phys. Rev. D} {\bf 80}
  (2009) 075018, [\href{http://arxiv.org/abs/0906.0580}{{\tt
  arXiv:0906.0580}}].

\bibitem{Dent:2012mx}
J.~B. Dent, F.~Ferrer, and L.~M. Krauss, {\it {Constraints on Light Hidden
  Sector Gauge Bosons from Supernova Cooling}},
  \href{http://arxiv.org/abs/1201.2683}{{\tt arXiv:1201.2683}}.

\bibitem{Rrapaj:2015wgs}
E.~Rrapaj and S.~Reddy, {\it {Nucleon-nucleon bremsstrahlung of dark gauge
  bosons and revised supernova constraints}},  {\em Phys. Rev. C} {\bf 94}
  (2016), no.~4 045805, [\href{http://arxiv.org/abs/1511.09136}{{\tt
  arXiv:1511.09136}}].

\bibitem{Dreiner:2013mua}
H.~K. Dreiner, J.-F. Fortin, C.~Hanhart, and L.~Ubaldi, {\it {Supernova
  constraints on MeV dark sectors from $e^+e^-$ annihilations}},  {\em Phys.
  Rev. D} {\bf 89} (2014), no.~10 105015,
  [\href{http://arxiv.org/abs/1310.3826}{{\tt arXiv:1310.3826}}].

\bibitem{Chang:2018rso}
J.~H. Chang, R.~Essig, and S.~D. McDermott, {\it {Supernova 1987A Constraints
  on Sub-GeV Dark Sectors, Millicharged Particles, the QCD Axion, and an
  Axion-like Particle}},  {\em JHEP} {\bf 09} (2018) 051,
  [\href{http://arxiv.org/abs/1803.00993}{{\tt arXiv:1803.00993}}].

\bibitem{Camalich:2020wac}
J.~M. Camalich, J.~Terol-Calvo, L.~Tolos, and R.~Ziegler, {\it {Supernova
  Constraints on Dark Flavored Sectors}},
  \href{http://arxiv.org/abs/2012.11632}{{\tt arXiv:2012.11632}}.

\bibitem{Kainulainen:1990bn}
K.~Kainulainen, J.~Maalampi, and J.~T. Peltoniemi, {\it {Inert neutrinos in
  supernovae}},  {\em Nucl. Phys. B} {\bf 358} (1991) 435--446.

\bibitem{Hanhart:2000er}
C.~Hanhart, D.~R. Phillips, S.~Reddy, and M.~J. Savage, {\it {Extra dimensions,
  SN1987a, and nucleon-nucleon scattering data}},  {\em Nucl. Phys. B} {\bf
  595} (2001) 335--359, [\href{http://arxiv.org/abs/nucl-th/0007016}{{\tt
  nucl-th/0007016}}].

\bibitem{Hanhart:2001fx}
C.~Hanhart, J.~A. Pons, D.~R. Phillips, and S.~Reddy, {\it {The Likelihood of
  GODs' existence: Improving the SN1987a constraint on the size of large
  compact dimensions}},  {\em Phys. Lett. B} {\bf 509} (2001) 1--9,
  [\href{http://arxiv.org/abs/astro-ph/0102063}{{\tt astro-ph/0102063}}].

\bibitem{Chang:2016ntp}
J.~H. Chang, R.~Essig, and S.~D. McDermott, {\it {Revisiting Supernova 1987A
  Constraints on Dark Photons}},  {\em JHEP} {\bf 01} (2017) 107,
  [\href{http://arxiv.org/abs/1611.03864}{{\tt arXiv:1611.03864}}].

\bibitem{Hardy:2016kme}
E.~Hardy and R.~Lasenby, {\it {Stellar cooling bounds on new light particles:
  plasma mixing effects}},  {\em JHEP} {\bf 02} (2017) 033,
  [\href{http://arxiv.org/abs/1611.05852}{{\tt arXiv:1611.05852}}].

\bibitem{Arvanitaki:2009fg}
A.~Arvanitaki, S.~Dimopoulos, S.~Dubovsky, N.~Kaloper, and J.~March-Russell,
  {\it {String Axiverse}},  {\em Phys. Rev. D} {\bf 81} (2010) 123530,
  [\href{http://arxiv.org/abs/0905.4720}{{\tt arXiv:0905.4720}}].

\bibitem{Peccei:1977np}
R.~D. Peccei and H.~R. Quinn, {\it {Some Aspects of Instantons}},  {\em Nuovo
  Cim. A} {\bf 41} (1977) 309.

\bibitem{Peccei:1977hh}
R.~D. Peccei and H.~R. Quinn, {\it {CP Conservation in the Presence of
  Instantons}},  {\em Phys. Rev. Lett.} {\bf 38} (1977) 1440--1443.

\bibitem{Weinberg:1977ma}
S.~Weinberg, {\it {A New Light Boson?}},  {\em Phys. Rev. Lett.} {\bf 40}
  (1978) 223--226.

\bibitem{Wilczek:1977pj}
F.~Wilczek, {\it {Problem of Strong $P$ and $T$ Invariance in the Presence of
  Instantons}},  {\em Phys. Rev. Lett.} {\bf 40} (1978) 279--282.

\bibitem{Abbott:1982af}
L.~F. Abbott and P.~Sikivie, {\it {A Cosmological Bound on the Invisible
  Axion}},  {\em Phys. Lett. B} {\bf 120} (1983) 133--136.

\bibitem{Dine:1982ah}
M.~Dine and W.~Fischler, {\it {The Not So Harmless Axion}},  {\em Phys. Lett.
  B} {\bf 120} (1983) 137--141.

\bibitem{Preskill:1982cy}
J.~Preskill, M.~B. Wise, and F.~Wilczek, {\it {Cosmology of the Invisible
  Axion}},  {\em Phys. Lett. B} {\bf 120} (1983) 127--132.

\bibitem{Nelson:2011sf}
A.~E. Nelson and J.~Scholtz, {\it {Dark Light, Dark Matter and the Misalignment
  Mechanism}},  {\em Phys. Rev. D} {\bf 84} (2011) 103501,
  [\href{http://arxiv.org/abs/1105.2812}{{\tt arXiv:1105.2812}}].

\bibitem{Arias:2012az}
P.~Arias, D.~Cadamuro, M.~Goodsell, J.~Jaeckel, J.~Redondo, and A.~Ringwald,
  {\it {WISPy Cold Dark Matter}},  {\em JCAP} {\bf 06} (2012) 013,
  [\href{http://arxiv.org/abs/1201.5902}{{\tt arXiv:1201.5902}}].

\bibitem{Battaglieri:2017aum}
M.~Battaglieri et~al., {\it {US Cosmic Visions: New Ideas in Dark Matter 2017:
  Community Report}},  in {\em {U.S. Cosmic Visions: New Ideas in Dark
  Matter}}, 7, 2017.
\newblock \href{http://arxiv.org/abs/1707.04591}{{\tt arXiv:1707.04591}}.

\bibitem{Gninenko:2001hx}
S.~N. Gninenko and N.~V. Krasnikov, {\it {The Muon anomalous magnetic moment
  and a new light gauge boson}},  {\em Phys. Lett. B} {\bf 513} (2001) 119,
  [\href{http://arxiv.org/abs/hep-ph/0102222}{{\tt hep-ph/0102222}}].

\bibitem{Pospelov:2008zw}
M.~Pospelov, {\it {Secluded U(1) below the weak scale}},  {\em Phys. Rev. D}
  {\bf 80} (2009) 095002, [\href{http://arxiv.org/abs/0811.1030}{{\tt
  arXiv:0811.1030}}].

\bibitem{TuckerSmith:2010ra}
D.~Tucker-Smith and I.~Yavin, {\it {Muonic hydrogen and MeV forces}},  {\em
  Phys. Rev. D} {\bf 83} (2011) 101702,
  [\href{http://arxiv.org/abs/1011.4922}{{\tt arXiv:1011.4922}}].

\bibitem{Batell:2011qq}
B.~Batell, D.~McKeen, and M.~Pospelov, {\it {New Parity-Violating Muonic Forces
  and the Proton Charge Radius}},  {\em Phys. Rev. Lett.} {\bf 107} (2011)
  011803, [\href{http://arxiv.org/abs/1103.0721}{{\tt arXiv:1103.0721}}].

\bibitem{Altmannshofer:2017bsz}
W.~Altmannshofer, M.~J. Baker, S.~Gori, R.~Harnik, M.~Pospelov, E.~Stamou, and
  A.~Thamm, {\it {Light resonances and the low-q$^{2}$ bin of $ {R}_{K^{*}}
  $}},  {\em JHEP} {\bf 03} (2018) 188,
  [\href{http://arxiv.org/abs/1711.07494}{{\tt arXiv:1711.07494}}].

\bibitem{Kaneta:2016wvf}
K.~Kaneta, H.-S. Lee, and S.~Yun, {\it {Portal Connecting Dark Photons and
  Axions}},  {\em Phys. Rev. Lett.} {\bf 118} (2017), no.~10 101802,
  [\href{http://arxiv.org/abs/1611.01466}{{\tt arXiv:1611.01466}}].

\bibitem{Kaneta:2017wfh}
K.~Kaneta, H.-S. Lee, and S.~Yun, {\it {Dark photon relic dark matter
  production through the dark axion portal}},  {\em Phys. Rev. D} {\bf 95}
  (2017), no.~11 115032, [\href{http://arxiv.org/abs/1704.07542}{{\tt
  arXiv:1704.07542}}].

\bibitem{Pospelov:2018kdh}
M.~Pospelov, J.~Pradler, J.~T. Ruderman, and A.~Urbano, {\it {Room for New
  Physics in the Rayleigh-Jeans Tail of the Cosmic Microwave Background}},
  {\em Phys. Rev. Lett.} {\bf 121} (2018), no.~3 031103,
  [\href{http://arxiv.org/abs/1803.07048}{{\tt arXiv:1803.07048}}].

\bibitem{Choi:2018mvk}
K.~Choi, S.~Lee, H.~Seong, and S.~Yun, {\it {Gamma-ray spectral modulations
  induced by photon-ALP-dark photon oscillations}},  {\em Phys. Rev. D} {\bf
  101} (2020), no.~4 043007, [\href{http://arxiv.org/abs/1806.09508}{{\tt
  arXiv:1806.09508}}].

\bibitem{Kalashev:2018bra}
O.~E. Kalashev, A.~Kusenko, and E.~Vitagliano, {\it {Cosmic infrared background
  excess from axionlike particles and implications for multimessenger
  observations of blazars}},  {\em Phys. Rev. D} {\bf 99} (2019), no.~2 023002,
  [\href{http://arxiv.org/abs/1808.05613}{{\tt arXiv:1808.05613}}].

\bibitem{Biswas:2019lcp}
S.~Biswas, A.~Chatterjee, E.~Gabrielli, and B.~Mele, {\it {Probing
  dark-axionlike particle portals at future $e^+e^-$ colliders}},  {\em Phys.
  Rev. D} {\bf 100} (2019), no.~11 115040,
  [\href{http://arxiv.org/abs/1906.10608}{{\tt arXiv:1906.10608}}].

\bibitem{Choi:2019jwx}
K.~Choi, H.~Seong, and S.~Yun, {\it {Axion-photon-dark photon oscillation and
  its implication for 21 cm observation}},  {\em Phys. Rev. D} {\bf 102}
  (2020), no.~7 075024, [\href{http://arxiv.org/abs/1911.00532}{{\tt
  arXiv:1911.00532}}].

\bibitem{Hook:2019hdk}
A.~Hook, G.~Marques-Tavares, and Y.~Tsai, {\it {Scalars Gliding through an
  Expanding Universe}},  {\em Phys. Rev. Lett.} {\bf 124} (2020), no.~21
  211801, [\href{http://arxiv.org/abs/1912.08817}{{\tt arXiv:1912.08817}}].

\bibitem{deNiverville:2020qoo}
P.~deNiverville, H.-S. Lee, and Y.-M. Lee, {\it {New searches at the reactor
  experiments based on the dark axion portal}},
  \href{http://arxiv.org/abs/2011.03276}{{\tt arXiv:2011.03276}}.

\bibitem{Arias:2020tzl}
P.~Arias, A.~Arza, J.~Jaeckel, and D.~Vargas-Arancibia, {\it {Hidden Photon
  Dark Matter Interacting via Axion-like Particles}},
  \href{http://arxiv.org/abs/2007.12585}{{\tt arXiv:2007.12585}}.

\bibitem{Graham:2015rva}
P.~W. Graham, J.~Mardon, and S.~Rajendran, {\it {Vector Dark Matter from
  Inflationary Fluctuations}},  {\em Phys. Rev. D} {\bf 93} (2016), no.~10
  103520, [\href{http://arxiv.org/abs/1504.02102}{{\tt arXiv:1504.02102}}].

\bibitem{Agrawal:2018vin}
P.~Agrawal, N.~Kitajima, M.~Reece, T.~Sekiguchi, and F.~Takahashi, {\it {Relic
  Abundance of Dark Photon Dark Matter}},  {\em Phys. Lett. B} {\bf 801} (2020)
  135136, [\href{http://arxiv.org/abs/1810.07188}{{\tt arXiv:1810.07188}}].

\bibitem{Bastero-Gil:2018uel}
M.~Bastero-Gil, J.~Santiago, L.~Ubaldi, and R.~Vega-Morales, {\it {Vector dark
  matter production at the end of inflation}},  {\em JCAP} {\bf 04} (2019) 015,
  [\href{http://arxiv.org/abs/1810.07208}{{\tt arXiv:1810.07208}}].

\bibitem{Co:2018lka}
R.~T. Co, A.~Pierce, Z.~Zhang, and Y.~Zhao, {\it {Dark Photon Dark Matter
  Produced by Axion Oscillations}},  {\em Phys. Rev. D} {\bf 99} (2019), no.~7
  075002, [\href{http://arxiv.org/abs/1810.07196}{{\tt arXiv:1810.07196}}].

\bibitem{Dror:2018pdh}
J.~A. Dror, K.~Harigaya, and V.~Narayan, {\it {Parametric Resonance Production
  of Ultralight Vector Dark Matter}},  {\em Phys. Rev. D} {\bf 99} (2019),
  no.~3 035036, [\href{http://arxiv.org/abs/1810.07195}{{\tt
  arXiv:1810.07195}}].

\bibitem{Long:2019lwl}
A.~J. Long and L.-T. Wang, {\it {Dark Photon Dark Matter from a Network of
  Cosmic Strings}},  {\em Phys. Rev. D} {\bf 99} (2019), no.~6 063529,
  [\href{http://arxiv.org/abs/1901.03312}{{\tt arXiv:1901.03312}}].

\bibitem{longpaper}
A.~Hook, G.~Marques-Tavares, C.~Ristow, and Y.~Tsai {in preparation}.

\bibitem{Braaten:1993jw}
E.~Braaten and D.~Segel, {\it {Neutrino energy loss from the plasma process at
  all temperatures and densities}},  {\em Phys. Rev. D} {\bf 48} (1993)
  1478--1491, [\href{http://arxiv.org/abs/hep-ph/9302213}{{\tt
  hep-ph/9302213}}].

\bibitem{Raffelt:2001kv}
G.~G. Raffelt, {\it {Muon-neutrino and tau-neutrino spectra formation in
  supernovae}},  {\em Astrophys. J.} {\bf 561} (2001) 890--914,
  [\href{http://arxiv.org/abs/astro-ph/0105250}{{\tt astro-ph/0105250}}].

\bibitem{DeRocco:2019jti}
W.~DeRocco, P.~W. Graham, D.~Kasen, G.~Marques-Tavares, and S.~Rajendran, {\it
  {Supernova signals of light dark matter}},  {\em Phys. Rev. D} {\bf 100}
  (2019), no.~7 075018, [\href{http://arxiv.org/abs/1905.09284}{{\tt
  arXiv:1905.09284}}].

\bibitem{Dreiner:2013tja}
H.~K. Dreiner, J.-F. Fortin, J.~Isern, and L.~Ubaldi, {\it {White Dwarfs
  constrain Dark Forces}},  {\em Phys. Rev. D} {\bf 88} (2013) 043517,
  [\href{http://arxiv.org/abs/1303.7232}{{\tt arXiv:1303.7232}}].

\bibitem{deNiverville:2018hrc}
P.~deNiverville, H.-S. Lee, and M.-S. Seo, {\it {Implications of the dark axion
  portal for the muon g-2, B-factories, fixed target neutrino experiments and
  beam dumps}},  {\em Phys. Rev. D} {\bf 98} (2018), no.~11 115011,
  [\href{http://arxiv.org/abs/1806.00757}{{\tt arXiv:1806.00757}}].

\bibitem{Capozzi:2020cbu}
F.~Capozzi and G.~Raffelt, {\it {Axion and neutrino bounds improved with new
  calibrations of the tip of the red-giant branch using geometric distance
  determinations}},  {\em Phys. Rev. D} {\bf 102} (2020), no.~8 083007,
  [\href{http://arxiv.org/abs/2007.03694}{{\tt arXiv:2007.03694}}].

\bibitem{Griest:1990kh}
K.~Griest and D.~Seckel, {\it {Three exceptions in the calculation of relic
  abundances}},  {\em Phys. Rev. D} {\bf 43} (1991) 3191--3203.

\bibitem{DAgnolo:2017dbv}
R.~T. D'Agnolo, D.~Pappadopulo, and J.~T. Ruderman, {\it {Fourth Exception in
  the Calculation of Relic Abundances}},  {\em Phys. Rev. Lett.} {\bf 119}
  (2017), no.~6 061102, [\href{http://arxiv.org/abs/1705.08450}{{\tt
  arXiv:1705.08450}}].

\bibitem{Fischer:2016cyd}
T.~Fischer, S.~Chakraborty, M.~Giannotti, A.~Mirizzi, A.~Payez, and
  A.~Ringwald, {\it {Probing axions with the neutrino signal from the next
  galactic supernova}},  {\em Phys. Rev. D} {\bf 94} (2016), no.~8 085012,
  [\href{http://arxiv.org/abs/1605.08780}{{\tt arXiv:1605.08780}}].

\end{thebibliography}\endgroup


\providecommand{\href}[2]{#2}\begingroup\raggedright\endgroup

\end{document}